\documentclass[10pt,twocolumn]{article}
\usepackage{times}

\usepackage{graphicx}
\usepackage{color}
\usepackage{balance}  
\usepackage{algorithm}
\usepackage[noend]{algpseudocode}
\usepackage{multirow}

\begin{document}

\algnewcommand{\LineComment}[1]{\Statex \(\triangleright\) #1}

\algrenewcommand{\algorithmicindent}{0.8em}

\newcommand{\sectionpadding}[0]{
}

\newtheorem{example}{Example}

\newcommand{\eat}[1]{}



\newcommand{\onecolfigures}[7]{
\begin{figure}[htb!]
	\scriptsize
	\centering
	\begin{minipage}{.49\linewidth}
	\centering
		\includegraphics[width=\linewidth]{#1}
		(a) Bayes
	\end{minipage}
	\begin{minipage}{.49\linewidth}
	\centering
		\includegraphics[width=\linewidth]{#2}
		(b) Wordcount
	\end{minipage}
	\begin{minipage}{.49\linewidth}
	\centering
		\includegraphics[width=\linewidth]{#3}
		(c) Media Streaming
	\end{minipage}
	\begin{minipage}{.49\linewidth}
	\centering
		\includegraphics[width=\linewidth]{#4}
		(d) MongoDB
	\end{minipage}
	\vspace{#7}
	\caption{#5}
	\label{#6}
\end{figure}
}

\newcommand{\twocolfigures}[6]{
\begin{figure*}[htb!]
	\scriptsize
	\centering
	\begin{minipage}{.245\linewidth}
	\centering
		\includegraphics[width=\linewidth]{#1}
		(a) Bayes
	\end{minipage}
	\begin{minipage}{.245\linewidth}
	\centering
		\includegraphics[width=\linewidth]{#2}
		(b) Wordcount
	\end{minipage}
	\begin{minipage}{.245\linewidth}
	\centering
		\includegraphics[width=\linewidth]{#3}
		(c) Media Streaming
	\end{minipage}
	\begin{minipage}{.245\linewidth}
	\centering
		\includegraphics[width=\linewidth]{#4}
		(d) MongoDB
	\end{minipage}
	\caption{#5}
	\label{#6}
\end{figure*}
}

\newcommand{\twocolfivefigures}[7]{
\begin{figure*}[htb!]
	\scriptsize
	\centering
	\begin{minipage}{.195\linewidth}
	\centering
		\includegraphics[width=\linewidth]{#1}
		(a) k-means
	\end{minipage}
	\begin{minipage}{.195\linewidth}
	\centering
		\includegraphics[width=\linewidth]{#2}
		(b) Bayes
	\end{minipage}
	\begin{minipage}{.195\linewidth}
	\centering
		\includegraphics[width=\linewidth]{#3}
		(c) Wordcount
	\end{minipage}
	\begin{minipage}{.195\linewidth}
	\centering
		\includegraphics[width=\linewidth]{#4}
		(d) Media Streaming
	\end{minipage}
	\begin{minipage}{.195\linewidth}
	\centering
		\includegraphics[width=\linewidth]{#5}
		(e) MongoDB
	\end{minipage}
	\vspace{-0.3cm}
	\caption{#6}
	\label{#7}
\end{figure*}
}

\newenvironment{myitemize}
{
    \begin{list}{\labelitemi}{\leftmargin=1em}
        \setlength{\topsep}{0pt}
        \setlength{\parskip}{0pt}
        \setlength{\partopsep}{0pt}
        \setlength{\parsep}{0pt}
        \setlength{\itemsep}{0pt}
}
{
    \end{list}
}

\title{A Decision Tree Based Approach Towards Adaptive Profiling of Distributed Applications (Extended Version)}
\author{
Ioannis Giannakopoulos$^1$, Dimitrios Tsoumakos$^2$ and Nectarios Koziris$^1$ \\
\begin{minipage}[htb]{.4\linewidth}		
		\centering
\small{$^1$ Computing Systems Laboratory}\\
\small{School of ECE, NTUA, Athens, Greece} \\
\small{\{ggian, nkoziris\}@cslab.ece.ntua.gr}
\end{minipage}
\quad
\begin{minipage}[htb]{.4\linewidth}
		\centering
\small{$^2$ Department of Informatics} \\
\small{Ionian University, Corfu, Greece} \\
\small{dtsouma@ionio.gr}
\end{minipage}
}
\date{}
\maketitle

\begin{abstract}
The adoption of the distributed paradigm has allowed applications to increase their scalability, robustness and fault tolerance, but it has also complicated their structure, leading to an exponential growth of the applications' configuration space and increased difficulty in predicting their performance. In this work, we describe a novel, automated profiling methodology that makes no assumptions on application structure. 
Our approach utilizes oblique Decision Trees in order to recursively partition an application's configuration space in disjoint regions, choose a set of representative samples from each subregion according to a defined policy and return a model for the entire space as a composition of linear models over each subregion. 
An extensive evaluation over real-life applications and synthetic performance functions showcases that our scheme outperforms other state-of-the-art profiling methodologies. 
It particularly excels at reflecting abnormalities and discontinuities of the performance function, allowing the user to influence the sampling policy based on the modeling accuracy and the space coverage.

\end{abstract}
\sectionpadding
\section{Introduction}
Performance modeling is a well-researched problem \cite{kapadia1999predictive,stewart2005performance,kerbyson2001predictive}. The identification of an application's behavior under different configurations is a key factor for it to be able to fulfill its objectives. As the application landscape evolves, new architectures and design patterns mature, enabling an increasing number of applications to be deployed in a distributed manner and benefit from the merits of this approach: Scalability, robustness and fault tolerance are some of the properties that render distributed platforms alluring and explain their wide adoption. By virtue of their design, distributed applications are commonly deployed to cloud infrastructures \cite{marston2011cloud}, in order to combine their inherent characteristics with the power of the cloud: Seemingly infinite resources, dynamically allocated and purchased, enable a distributed application to scale in a cost effective way. However, the adoption of the distributed paradigm increased the complexity of the application architecture, for two reasons: First, many assisting software modules that support coordination, cluster management, etc., are essential for the application to run properly. Second, each module can be configured in numerous ways and the configurations are usually independent of each other. For these reasons, the application configuration space has vastly expanded and, hence, the problem of modeling an application performance, also called an \emph{application profile}, has become particularly complicated.

Evidently, an automated estimation of a distributed application profile would prove highly beneficial. The magnitude of the configuration space renders exhaustive approaches that demand the exploration of a massive part of the configuration space impractical, since they entail both a prohibitive number of deployments and an enormous amount of computation. Several approaches target to model the application performance \cite{fittkau2012cdosim,wickremasinghe2010cloudanalyst,li2010webprophet,mytilinis2015performance} in an analytical way. These approaches are effective for known applications with specific structure and they are based on simulation or emulation techniques \cite{mesnier2007trace,hoste2006performance}. To overcome the rigidness of these schemes, other methods \cite{gonccalves2015performance,giannakopoulos2015panic,kundu2010application,kundu2012modeling} take a ``black-box'' approach, in which the application receives a set of inputs, corresponding to the different factors of the configuration space and produces one or more outputs, corresponding to its performance. These approaches try to identify the relationship between the input and the output variables for a subset of the configuration space (utilizing sampling) and generalize the findings with Machine Learning techniques (modeling). 

\begin{figure}[htb!]
		\centering
		\includegraphics[width=.45\linewidth]{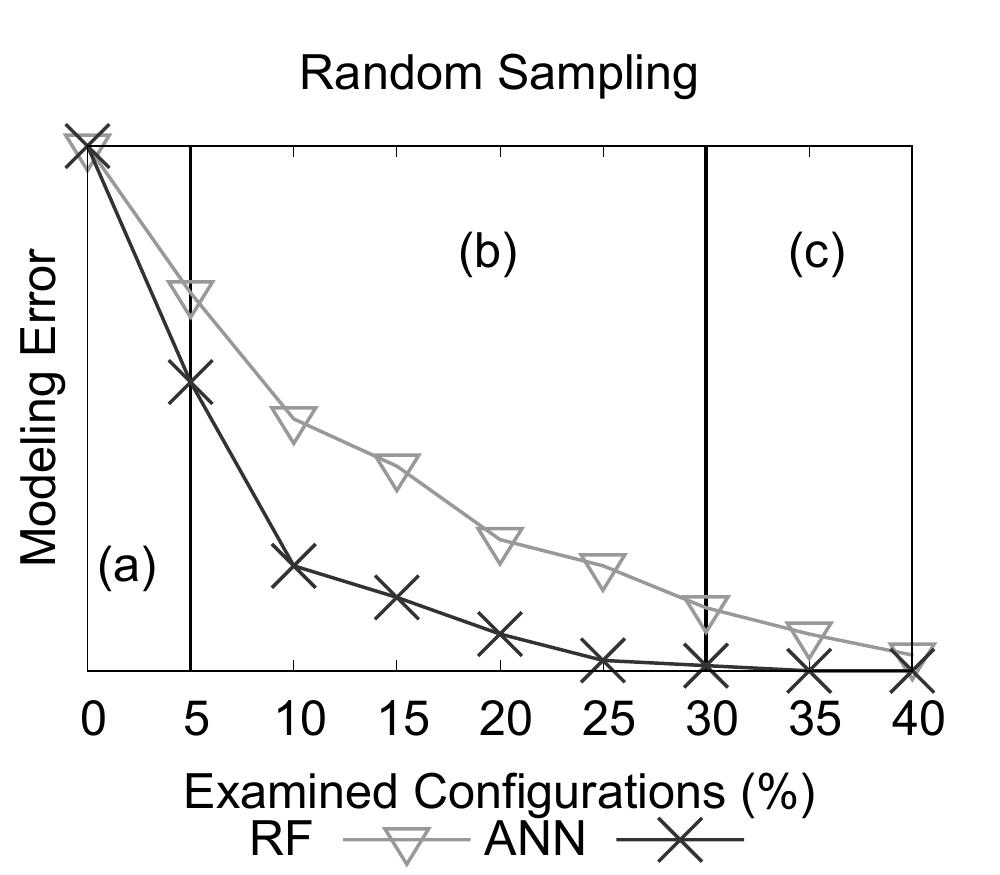}
		\includegraphics[width=.45\linewidth]{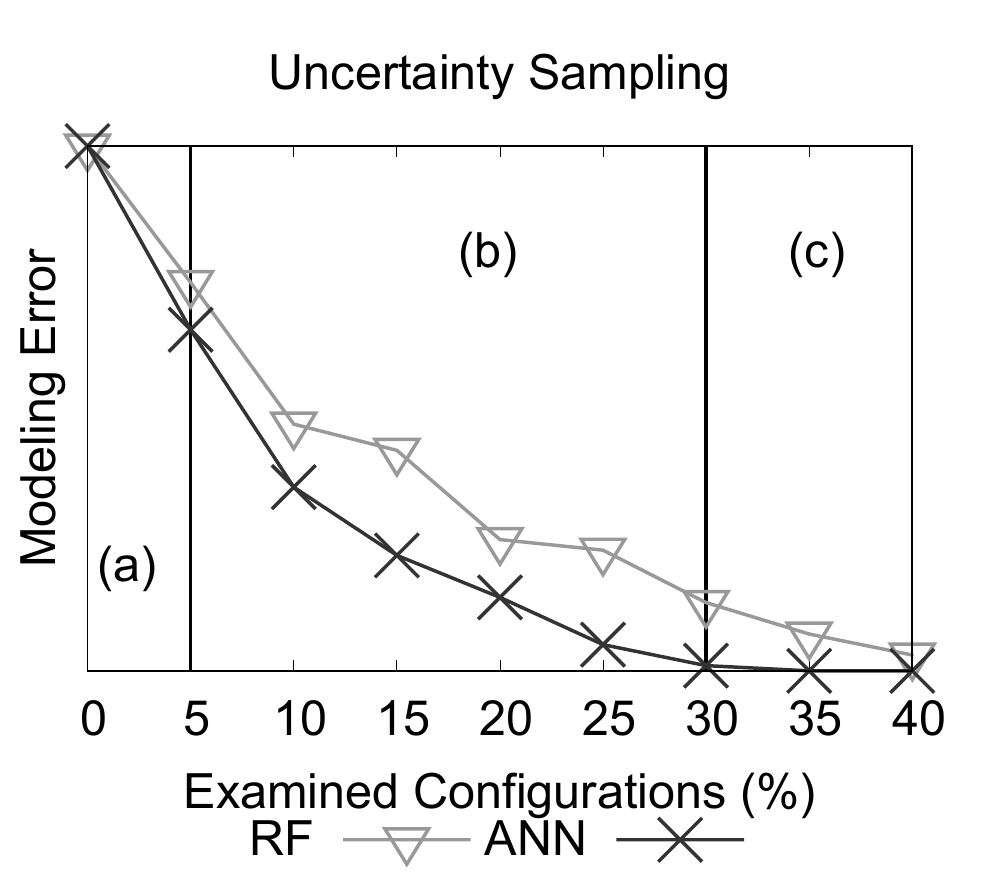}
		\vspace{-0.3cm}
		\caption{Profiling errors for random and optimal schemes}
		\label{fig:example}
\end{figure}

Such profiling approaches require multiple application deployments for distinct configurations in order to efficiently explore the configuration space and capture its behavior: The larger the portion of the configuration space examined, the more accurate the profiling. The choice of the configurations, though, is a decisive factor: For a given number of deployments, the more representative the chosen configurations, the more accurate the profile. In Figure \ref{fig:example}, we provide the error for predicting the execution time of the Wordcount operator for a given dataset when launched over Hadoop clusters of different sizes, using two different classifiers (Random Forests and an Artificial Neural Network) and two different sampling schemes (Random, depicted in the left figure and Uncertainty Sampling \cite{lewis1994heterogeneous} depicted in the right). The horizontal axis represents the number of configurations which are tested for each run, expressed as a portion of the total number of configurations: The larger the percentage, the more configurations are considered. Generally speaking, three areas can be identified by the two plots: Area (a), when a short number of configurations is examined, the modeling error is high and independent of the sampling and modeling schemes, since much more information is required in order to model the application. On the contrary, area (c) represents the case where multiple configurations are already available. Albeit the modeling error still slightly diminishes for both classifiers, a large number of application deployments has already taken place. Finally, area (b) represents the intermediate case, as the deployment of more configurations leads to a radical error degradation but the number of deployment is not yet prohibitively costly. Obviously, regardless of the sampling and the modeling schemes, region (b) is the most important one, since it combines a reasonable number of deployments with modeling of satisfying accuracy.




\vspace{-0.4cm}
\vspace{-0.4cm}

In this work, we propose an adaptive approach to automatically generate an application profile for any distributed application given a specific number of deployments, particularly focusing on appropriately selecting the tested configurations in order to maximize the modeling accuracy. Our work tackles the sampling and modeling steps in a unified way: First, we introduce an accuracy-driven sampling technique that favors regions of the configuration space which are not accurately approximated. Second, we decompose the configuration space in disjoint regions and utilize different models to approximate the application performance for each of them. 
The basis of our approach lies on the mechanics of Classification and Regression Trees \cite{breiman1984classification} and relies on recursively partitioning the configuration space of the application into disjoint regions. Each partition is then assigned with a number of configurations to be deployed, with the process being iterated until a pre-defined maximum number of sample configurations is reached. The number of configurations allocated at each region is adaptively decided according to the approximation error and the size of each partition. Finally, the entire space is approximated by linear models per partition, based on the deployed configurations. Intuitively, our approach attempts to ``zoom-in'' to regions where application performance is not accurately approximated, paying specific interest to all the abnormalities and discontinuities of the performance function. By utilizing oblique Decision Trees \cite{barros2011bottom}, we are able to capture patterns that are affected by multiple configuration parameters simultaneously. 
In summary, we make the following contributions:
\\
%
(a) We propose an adaptive, accuracy-driven profiling technique for distributed applications that utilizes oblique Decision Trees. Our method natively decomposes the multi-dimensional input space into disjoint regions, naturally adapting to the complex performance application behavior in a fully automated manner. 
Our scheme utilizes three unique features relative to the standard Decision Tree algorithm: First, it proposes a novel expansion algorithm that constructs oblique Decision Trees by examining whether the obtained samples fit into a linear model.
Second, it allows developers to provide a compromise between \emph{exploring} the configuration space and \emph{exploiting} the previously obtained knowledge. Third, it adaptively selects the most accurate modeling scheme, based on the achieved accuracy.\\
(b) We perform an extensive experimental evaluation over diverse, real-world applications and synthetic performance functions of various complexities. Our results showcase that our methodology is the most efficient, achieving modeling accuracies even $3\times$ higher that its competitors and, at the same time, it is able to create models that reflect abnormalities and discontinuities of the performance function orders of magnitude more accurately.
Furthermore, it is interesting to note that our sampling methodology proves to be particularly beneficial for linear classifiers and ensembles of them, as linear models trained with samples chosen by our scheme present even $38\%$ lower modeling error.

\vspace{-.1cm}
\sectionpadding
\section{Background}

In this section we provide the problem formulation and information relative to the background of our work.

\subsection{Problem formulation}
\label{section:background:problem}

The problem of creating a performance model for a distributed application can be formulated as a function approximation problem \cite{cunha2016cloud,giannakopoulos2015panic}. The application is viewed as a black-box that receives a number of inputs and produces a single (or more) output(s). The main idea behind constructing the performance model is to predict the relationship between the inputs and the output, without making any assumption regarding the application's architecture. In other words, one would want to identify a function that projects the input variables into the output for every possible input value. Inputs reflect any parameters that affect application performance: Number and quality of different types of resources (e.g., cores/memory, number of nodes for distributed applications, etc.), application-level parameters (e.g., cache used by an RDBMS, HDFS block size, replication factor, etc.), workload-specific parameters (e.g., throughput/type of requests) and dataset-specific parameters (e.g., size, dimensionality, distribution, etc.) are some of those.

Assume that an application comprises $n$ inputs and a single output. We assume that the $i^{th}$ input, $1\leq i \leq n$, may receive values from a predefined finite set of values, denoted as $d_i$. The Cartesian product of all $d_i, 1\leq i \leq n$ is referred to as the \emph{Deployment Space} of the application $D= d_1 \times d_2 \times \cdots \times d_n$. Similarly, the output of the application produces values that correspond to a performance metric, indicative of the application's ability to fulfill its objectives. The set of the application's output will be referred to as the \emph{Performance Space} of the application $P$. 
Based on the definitions of $D$ and $P$, we define the performance model $m$ of an application as a function mapping points from $D$ to $P$, i.e., $m : D \rightarrow P$. The estimation of the performance model of an application entails the estimation of the performance value $b_i \in P$ for each $a_i \in D$. However, $|D|$ increases exponentially with $n$ (since $ |D| = \prod_{i=1}^{n} |d_{i}|$), thus the identification of all the performance values becomes impractical, both in terms of computation and budget. A common approach to address this challenge is the extraction of a subset $D_s \subseteq D$, much smaller that the original Deployment Space and the estimation of the performance points $P_s$ for each $a_i \in D_s$. Using $D_s$ and $P_s$, model $m$ can  be approximated, creating an approximate model denoted as $m'$. 
The difference between $m$ and $m'$ is indicative of the accuracy of the later: The smaller the difference, the more accurate the approximate model becomes. The accuracy of $m'$ is greatly influenced by two factors: (a) the set $D_s$ and (b) the modeling method used to create $m'$. 
It should be emphasized that the application profiling problem resembles the \emph{Global Optimization} problem \cite{storn1997differential}, in which the optimal configuration for an application in order to achieve the highest performance is explored. The two problems, although sharing similar formulations, they greatly differ in the following: The later formulation seeks for an optimal point, whereas the former seeks for a suitable \emph{set} of points in order to construct an accurate model. 


Finally, we note that the previous formulation is valid only under the assumption that distinct application deployments are \emph{reproducible}, in the sense that in case of re-deployment of a given Deployment Space point, the measured outcome should be identical. For many reasons, such as the ``noisy neighbor'' effect \cite{garcia2014challenges}, network glitches, power outages, etc., such an assumption can be violated in cloud environments because of the introduced unpredictability that distorts the application's behavior. The treatment of this dimension of the problem is outside the scope of our work. This works tackles the complexity introduced by the excessive dimensionality of the Deployment Space. The presented methodology can be, thus, directly applied to predictable environments with reduced interference, such as private cloud installations that, according to \cite{rightscale}, remain extremely popular, since they will host half of the user generated workloads for 2017, maintaining the trend from the previous years.




\subsection{Decision Trees}
\label{section:background:dt}
\emph{Classification and Regression Trees (CART)} \cite{breiman1984classification}, or Decision Trees, are a very popular classification and regression approach. They are formed as tree structures containing a set of intermediate nodes (the \emph{test} nodes) and a set of \emph{leaf} nodes. Each test node represents a boundary of the space of the data and each leaf node represents a class, if the Decision Tree is used for classification, or a linear model, if it is used for regression. The boundaries of the Decision Tree divide the original space into a set of disjoint regions. 
The construction of a Decision Tree is based on recursively partitioning the space of the data so as to create disjoint groups of points that maximize their intra-group homogeneity and minimize their inter-group homogeneity. The homogeneity metric of a group differs among the existing algorithms: \emph{GINI impurity} has been used by the CART algorithm \cite{breiman1984classification}, \emph{Information Gain} has been used by ID3 \cite{quinlan1986induction} and C4.5 \cite{quinlan1993c4} for classification, whereas the Variance Reduction \cite{breiman1984classification} is commonly used for regression. The aforementioned heuristics are applied to each leaf to decide which dimension should be used for partitioning and at which value. 
The termination condition for the construction varies between different algorithms: In many cases, the tree length is pre-defined, whereas, in other cases, the termination condition is dictated by its accuracy (if the expansion of the tree marginally benefits its accuracy, the construction stops).

\begin{figure}[htb]
	\centering
	\hfill
	\includegraphics[width=.35\linewidth]{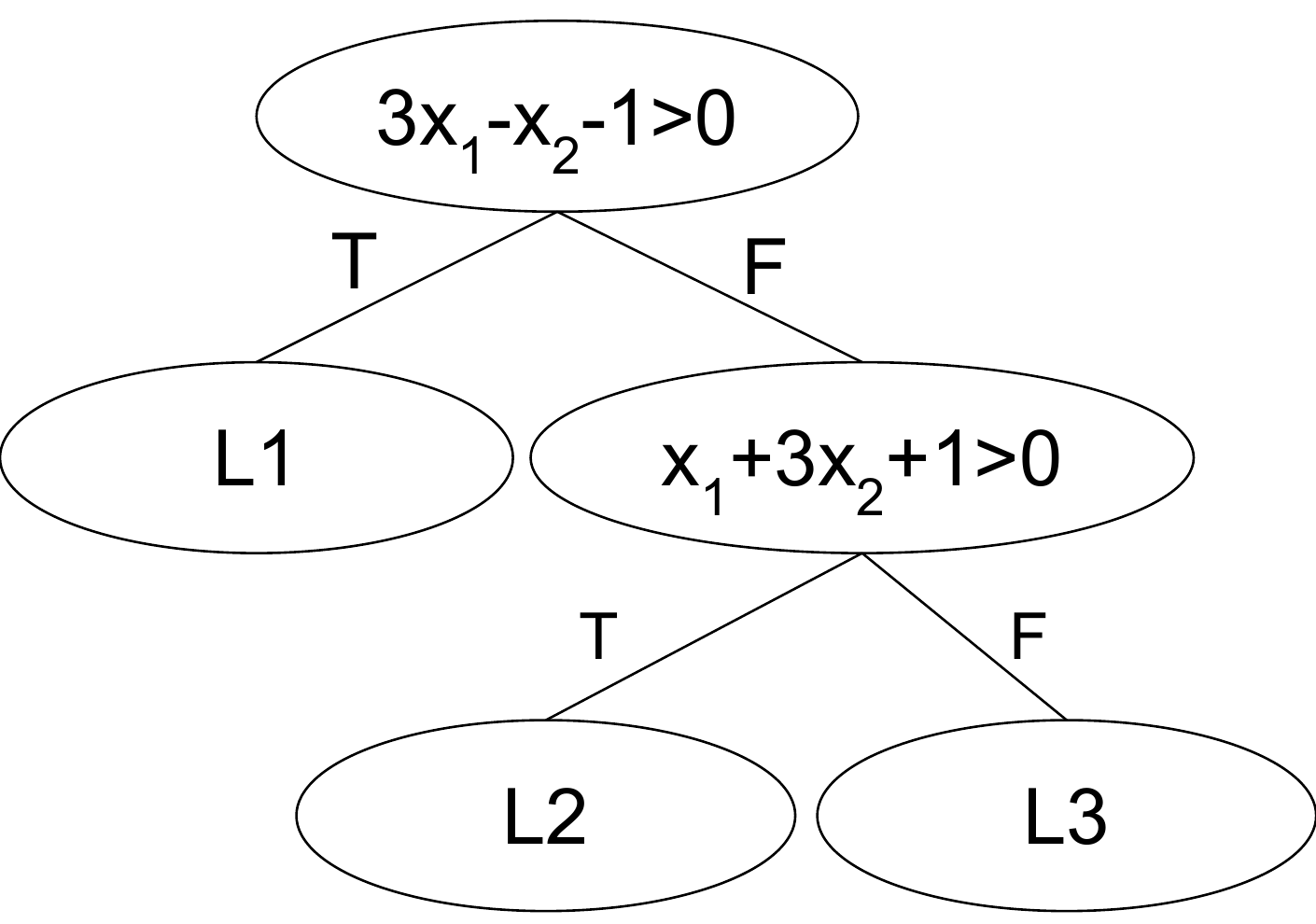}
	\hfill
	\includegraphics[width=.35\linewidth]{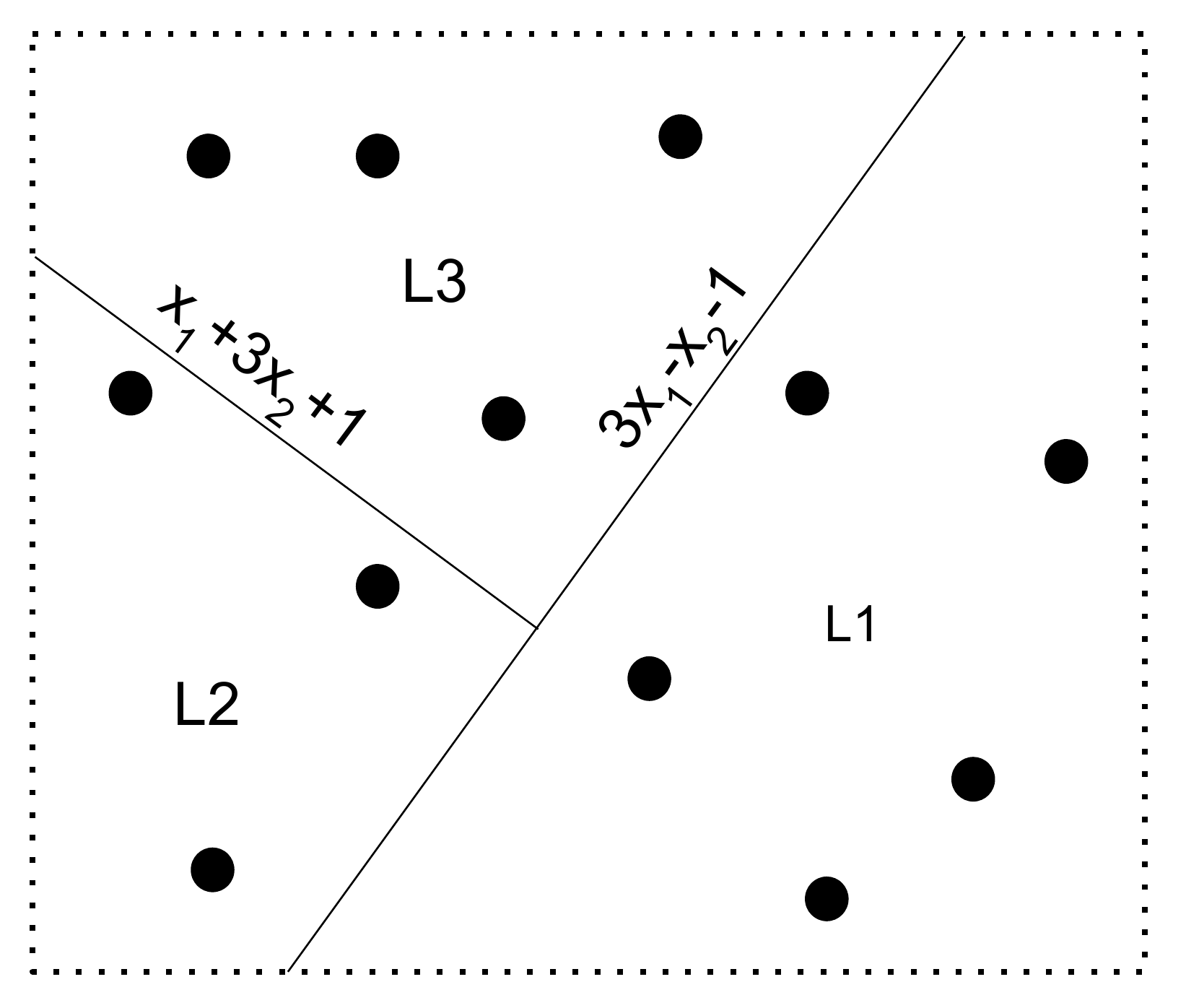}
	\hfill
	\vspace{-.3cm}
	\caption{Example of an Oblique Decision Tree}
	\label{fig:dt:example}
\end{figure}

Each boundary of a Decision Tree is parallel to one axis of the data, since it only involves a single dimension of the data space, i.e., the boundary line is expressed by a rule of the form $x_i = c$, where $c$ is a constant value. Generalizing this rule into a multivariate line, we obtain the \emph{oblique} Decision Trees \cite{barros2011bottom} that comprise of lines of the form:
$\sum_{i=1}^{n} c_i x_i + \gamma = 0$.
To better illustrate this, in Figure \ref{fig:dt:example} we provide an example where we showcase the tree structure along with the respective partitions of the Deployment Space for a tree that has 3 leaves and 2 test nodes. Original (or {\it flat}) Decision Trees can be considered as a special case of the oblique Decision Trees. The multivariate boundaries boost the expressiveness of the Decision Tree, since non axis-parallel patterns can be recognized and expressed, as discussed in \cite{heath1993induction}.

\vspace{-.5cm}
\vspace{-.3cm}

In this paper, we are utilizing oblique Decision Trees in two ways: First, we are employing them to create an approximate model of the performance function. Second, we are exploiting their construction algorithm to adaptively sample the Deployment Space of the application, focusing more on regions where the application presents a complex behavior and ignoring regions where its behavior tends to be predictable. These uses are extensively described in Sections \ref{section:profiling:construction} and \ref{section:profiling:sampling} respectively. 

\subsection{Method overview}
\vspace{-.5cm}
\begin{figure}[htb]
	\centering
	\includegraphics[width=0.85\linewidth]{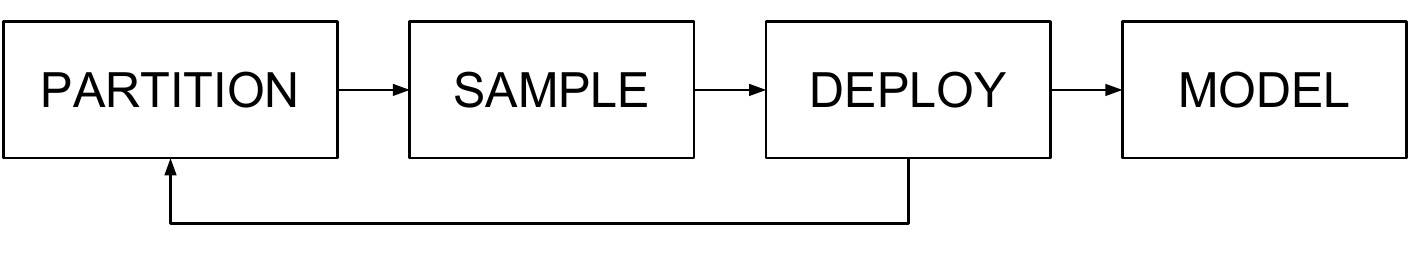}
	\vspace{-.4cm}
	\caption{Method overview}
	\label{fig:overview}
\end{figure}
\vspace{-.4cm}

The main idea of the suggested algorithm is to partition the Deployment Space of the application by grouping samples that better fit different linear models, obtain knowledge about the performance function through sampling the Deployment Space, deploying the selected configurations and, when a predefined number of deployments is reached, model the performance function utilizing different linear models for each partition. This is schematically represented in Figure \ref{fig:overview}. Specifically, at each step, the Deployment Space is partitioned by grouping the already obtained samples according to their ability to create a linear model that accurately approximates the application performance; Estimates are then created regarding the intra-group homogeneity. The homogeneity of each region corresponds to the accuracy of the prediction of the performance function for the specified region. Therefore, less accurate regions must be further sampled to clarify the behavior of the performance function in them. On the contrary, more accurate regions are considered successfully approximated, thus new samples need not be obtained for those. 

Intuitively, the suggested algorithm is an attempt to adaptively ``zoom-in'' to areas of the Deployment Space where the behavior of the performance function is more \emph{obscure}, in the sense that it is unpredictable and hard to model. This enables {\color{black}the number of allowed application deployments to be} dynamically distributed inside the Deployment Space, leading to more accurate predictions as more samples are collected for performance areas that are harder to approximate. The properties of Decision Trees, heavily utilized throughout this work, render them a perfect candidate for our approach: Their ease of implementation, scalability, robustness and, especially, their inherent divide-and-conquer functionality, that allows for adaptiveness in focusing on the regions of unpredictable application performance, are some of those.  These properties inherently match the application profiling problem and facilitate the decomposition of the Deployment Space in an intuitive and efficient manner.
\sectionpadding
\section{Profiling Methodology}
\label{section:profiling}
In this section we provide an extensive description of the profiling methodology. 

\subsection{Algorithm Overview}
\label{section:profiling:overview}
\vspace{-.3cm}
\begin{algorithm}
		\footnotesize
\caption{DT-based Adaptive Profiling Algorithm}
\label{algorithm:dtadaptive}
\begin{algorithmic}[1]
\Procedure{DTAdaptive}{$D$, $B$, $b$}
\State{$tree \leftarrow \Call{treeInit}{\emptyset},samples \leftarrow \emptyset$} 
\While{$|samples| \leq B$}
\State $tree \leftarrow $ \Call{expandTree}{$tree$, $samples$}
\State{$s \leftarrow $ \Call{sample}{$D$, $tree$, $samples$, $b$}}
\State{$d \leftarrow $ \Call{deploy}{$s$}}
\State{$samples \leftarrow samples \cup d$}
\EndWhile
\State{$model \leftarrow $ \Call{createModel}{$samples$}}
\State \Return{$model$}
\EndProcedure
\end{algorithmic}
\end{algorithm}
\vspace{-.3cm}

{\color{black}Our Algorithm takes three input parameters: (a) the Deployment Space of the application $D$, (b) the maximum number of application deployments $B$ and (c) the number of deployments triggered at each iteration of the algorithm $b$}. The \emph{tree} variable represents a Decision Tree, while the \emph{samples} set contains the obtained samples. While the number of obtained samples has not reached $B$, the following steps are executed: First, the leaves of the tree are examined and tested whether they can be replaced by subtrees that further partition their regions (Line 4). Next, the leaves of the expanded tree are sampled according to the \emph{SAMPLE} function (Line 5), and the chosen samples are deployed (Line 6) according to the deployment configuration of the sample, and performance metrics are obtained. One must note the difference between $s$ and $d$: $s \subseteq D$ whereas $d \subseteq D \times P$, where $D$ and $P$ are the Deployment and Performance space respectively, as defined in Section \ref{section:background:problem}. $s$ members are points of the Deployment Space representing a deployment configuration, whereas $d$ members represent a point of the performance function that \emph{contains} the respective input space point plus the performance value. {\color{black}Finally, when $B$ samples have been chosen, the final model is created (Line 8).}

Before describing each function in more detail, some observations can be made regarding the algorithm's execution. First, the tree is incrementally expanded at each iteration and its height is gradually increased. The arrival of new samples leads to the formulation of new rules for partitioning the Deployment Space (represented by test nodes in the tree structure), leading to an \emph{online} training scheme. Second, the \emph{DEPLOY} function (Line 6) is responsible for making $|s|$ deployments of the application according to the configuration setups specified by each $s_i \in s$ and return the same configuration points accompanied by their performance metric. This implies that $|s|=|d|$ and this is the most time consuming part of each iteration, since resource orchestration and application configuration typically requires several minutes. Lastly, it is obvious the total number of iterations is equal to $\lceil \frac{B}{b} \rceil$.

{\color{black}
\subsection{Decision Tree Expansion}
\label{section:profiling:construction}

The main idea behind the expansion of a Decision Tree lies on the identification of a split line that maximizes the homogeneity between the two newly created leaves; The selection of an appropriate split line greatly impacts the accuracy of the partitioning, ergo the future partitions. As mentioned before, the utilization of \emph{oblique} Decision Trees allows the test node to be represented by a multivariate line and enhances their adaptability to different performance functions. Nevertheless, calculating an appropriate multivariate split line is NP-complete \cite{heath1993induction} and, thus, the optimal solution can only be approximated. Since we want to exploit the expressiveness of the oblique partitions without introducing a prohibitive computation cost for their estimation, we express the problem as an optimization problem \cite{heath1993induction} and utilize Simulated Annealing (SA) \cite{van1987simulated} to identify a near-optimal line. In Algorithm \ref{algorithm:treeexpansion}, we present the \texttt{EXPANDTREE} function. For each leaf of the tree (Line 3), SA is executed to identify the best multivariate split line (Line 5), considering solely the samples whose Deployment Space-related dimensions ($d.in$) lie inside the specified leaf (Line 4). The samples of the specified leaf are then partitioned in two disjoint sets according to their position (Lines 7-11), in which the symbol $\preceq$ indicates that a sample $s$ is located below $line$). Finally, a new test node is generated (Line 12), replacing the original leaf node of the tree and the new tree is returned.


SA is a prominent methodology for solving complex optimization problems.
It functions in an iterative manner and is based on the idea of generating possible solutions (split lines), evaluating their goodness using a score function and, finally, returning the best. The key factor that diversifies SA from a simple random search lies on the targeted creation of candidate solutions. Assuming a split line of the form  $l = a_1 x_1 + \cdots + a_n x_n + a_{n+1} = 0$ , SA seeks for a solution vector $v = (a_1, \cdots, a_{n+1})$ that minimizes an expression $Score(v)$; During the first algorithm steps ($i$), the consecutive candidate solutions present great differences , i.e., the difference $|v_i-v_{i+1}|$, is large, but for an increasing number of iterations, $|v_i-v_{i+1}|\rightarrow 0$ and SA converges to a close-to-the-optimal solution. During the algorithm's execution, SA may pick worse solutions, i.e., split lines with worse scores, as the best examined this-far; This guarantees that the algorithm will not be trapped into a local minimum. Both the probability of substituting a better solution with a worse one and the size of the neighborhood of solutions that SA examines at each step is determined by \emph{Temperature} factor, decaying with time. Intuitively, higher temperatures mean that SA attempts diverse solutions and accepts worse solutions with high probability and lower temperature implies that SA converges to the neighborhood of the best solution. 

\vspace{-.3cm}
\begin{algorithm}[htb!]
\footnotesize
\caption{EXPANDTREE Function}
\label{algorithm:treeexpansion}
\begin{algorithmic}[1]
\Procedure{expandTree}{$tree$, $samples$}
\State{$newTree \leftarrow tree$}
\For{$l \in$ leaves($tree$)}
\State{$v \leftarrow \{ d| d \in samples, d.in \in l \}$}
\State{$line \leftarrow SA(v)$}
\State{$L_1 \leftarrow \emptyset, L_2 \leftarrow \emptyset$}
\For{$s \in v$}
\If{$s \preceq line$}
\State{$L_1 \leftarrow L_1 \cup s$}
\Else{}
\State{$L_2 \leftarrow L_2 \cup s$}
\EndIf
\EndFor
\State{$testNode \leftarrow \{line$, $L1$, $L2$\}}
\State{$newTree \leftarrow $replace($l$, $testNode$)}
\EndFor{}
\State{\Return{$newTree$}}
\EndProcedure
\end{algorithmic}
\end{algorithm}
\vspace{-.3cm}

As easily understood, the cornerstone for SA's effectiveness is the score function that quantifies the efficacy of each candidate solution. The methodologies utilized by various Decision Tree construction algorithms make no assumption regarding the nature of the data; Although this enhances their adaptability to different problem spaces, their utilization in this work resulted in poorly partitioned Deployment Spaces. The data that our work attempts to model belong to a performance function. Intuitively, if all the performance function points were available, one would anticipate that a closer observation of a specific region of the Deployment Space would present an approximately linear behavior, as ``neighboring'' Deployment Space points are anticipated to produce similar performance. When focusing on a single neighborhood, this ``similar performance'' could be summarized by a linear hyperplane. A split line is considered to be ``good'' when the generated leaves are best summarized by linear regression models or, equivalently, if the samples located in the two leaves can produce linear models with low residuals, i.e., low modeling error. Given an $n$-dimensional Deployment Space $D$, a line $l=0$ and two sets $L_1$, $L_2$ that contain $D$'s samples after partitioning it with $l$ (as in lines 6-10 of the algorithm), we estimate two linear regression models for $L_1$ and $L_2$ and estimate their residuals using the coefficient of determination $R^2$; $l$'s score is:
$
Score(l) = -\frac{|L_1| R^2_{L_1} + |L_2| R^2_{L_2}}{|L_1| + |L_2|}
$.
Note that $0 \leq R^2 \leq 1$ and a value of $1$ represents zero residuals and perfect fit to the linear model while a value of $0$ implies the unsuitability of a linear model. The above score function is minimized when both $L_1$ and $L_2$ generate highly accurate linear models or, equivalently, if the two sets can be accurately represented by two linear hyperplanes. Notice that we weight their importance according to the number of samples they contain as an inaccurate linear model generated by more samples has greater impact to the algorithm than an inaccurate model generated by fewer samples. The negative sign is employed in order to remain aligned with the literature, in which SA seeks for minimum points. Finally, one SA property not discussed this far is the ability to achieve a customizable compromise between the modeling accuracy and the required computation. Specifically, prior to SA's execution, the user defines a maximum number of iterations to be executed for the identification of the best split line. When one wants to maximize the quality of the partitioning, many iterations are conducted. On the contrary, the number of iterations is set to lower values in order to allow for a quick estimation of the results.

}

\subsection{Adaptive Sampling}
\vspace{-.3cm}
\begin{algorithm}[htb!]
		\footnotesize
\caption{Sampling algorithm}
\label{algorithm:sampling}
\begin{algorithmic}[1]
\Procedure{sample}{$D$, $tree$, $samples$, $b$}
\State{$errors,sizes \leftarrow \emptyset$, $maxError,maxSize \leftarrow 0$}
\For{$l \in $ leaves($tree$) }
\State{$points \leftarrow \{d | d \in samples,  d.in \in l\}$}
\State{$m \leftarrow$ regression($points$)}
\State{$errors[l] \leftarrow$ crossValidation($m$,$points$)}
\State{$sizes[l] \leftarrow |\{e | e \in D \cap l\}|$}
\If{$maxError \leq errors[l]$}
\State{$maxError \leftarrow errors[l]$}
\EndIf
\If{$maxSize \leq sizes[l]$}
\State{$maxSize \leftarrow sizes[l]$}
\EndIf
\EndFor
\State{$scores, newSamples\leftarrow \emptyset$, $sumScores \leftarrow 0$  }
\For{$l \in $ leaves($tree$) }
\State{$scores[l] \leftarrow w_{error} \cdot \frac{errors[l]}{maxError} + w_{size} \cdot \frac{sizes[l]}{maxSize}$ }
\State{$sumScores \leftarrow sumScores + scores[l]$ }
\EndFor
\For{$l \in $ leaves($tree$) }
\State{$leafNoDeps\leftarrow  \lceil \frac{scores[l]}{sumScores} \cdot b \rceil$ }
\State{$s \leftarrow $\Call{randomSelect}{$\{ d | d \in D \cap l\}$, $leafNoDeps$}}
\State{$newSamples \leftarrow newSamples \cup s$}
\EndFor
\State{\Return $newSamples$}
\EndProcedure
\end{algorithmic}
\end{algorithm}
\vspace{-.3cm}

\label{section:profiling:sampling}
After the tree expansion, the \texttt{SAMPLE} function is executed (Algorithm \ref{algorithm:sampling}). 
The algorithm iterates over the leaves of the tree (Line 3); For the samples of each leaf, a new linear regression model is calculated (Line 5) and its residuals are estimated using Cross Validation \cite{geisser1993predictive}: The higher the residuals, the worse the fit of the points to the linear model. The size of the specified leaf is then estimated.  
After storing both the error and the size of each leaf into a dictionary, the maximum leaf error and size are calculated (Lines 8-11). Subsequently, a score is estimated for each leaf (Lines 13-15). The score of each leaf is set to be proportional to its scaled size and error. {\color{black} This normalization is conducted so as to guarantee that the impact of the two factors is equivalent.} Two coefficients $w_{error}$ and $w_{size}$ are used to assign different weights to each measure. These scores are accumulated and used to proportionally distribute $b$ to each leaf (Lines 16-19). In that loop, the number of deployments of the specified leaf is calculated and new samples from the subregion of the Deployment Space are randomly drawn with the \emph{RANDOMSELECT} function, in a uniform manner. 
Finally, the new samples set is returned. 

\vspace{-.3cm}
\begin{figure}[htb]
	\begin{minipage}{0.49\linewidth}
	\centering
	\includegraphics[width=\linewidth]{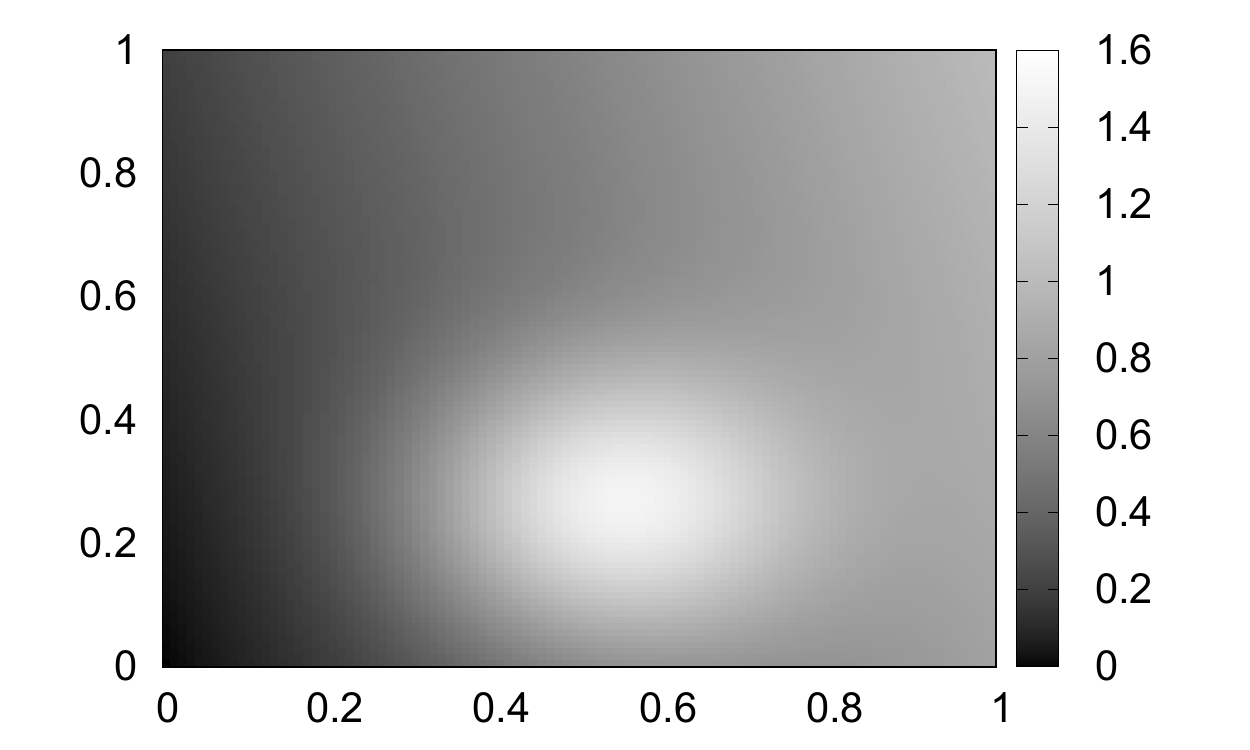}
	\scriptsize
	(a) Original function
	\end{minipage}
	\begin{minipage}{0.49\linewidth}
	\centering
	\includegraphics[width=\linewidth]{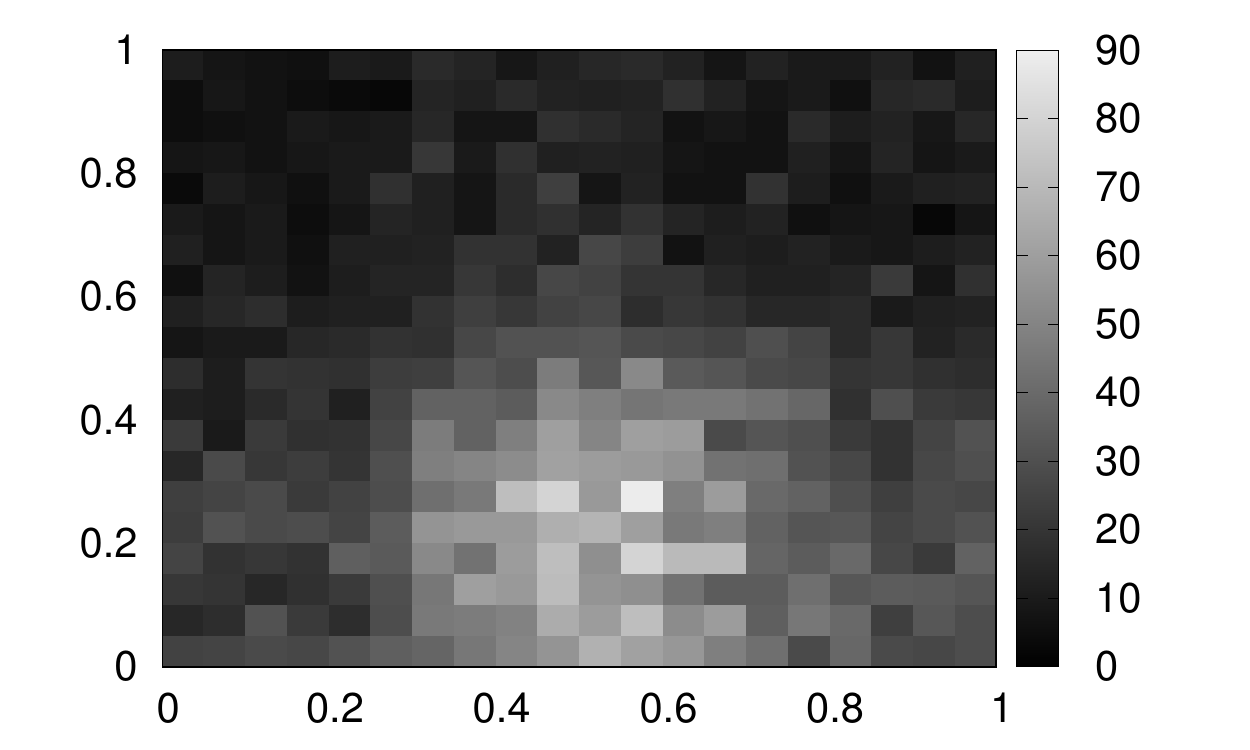}
	\scriptsize
	(b) $w_{error}=1.0, w_{size}=0.0$
	\end{minipage}
	\begin{minipage}{0.49\linewidth}
	\centering
	\includegraphics[width=\linewidth]{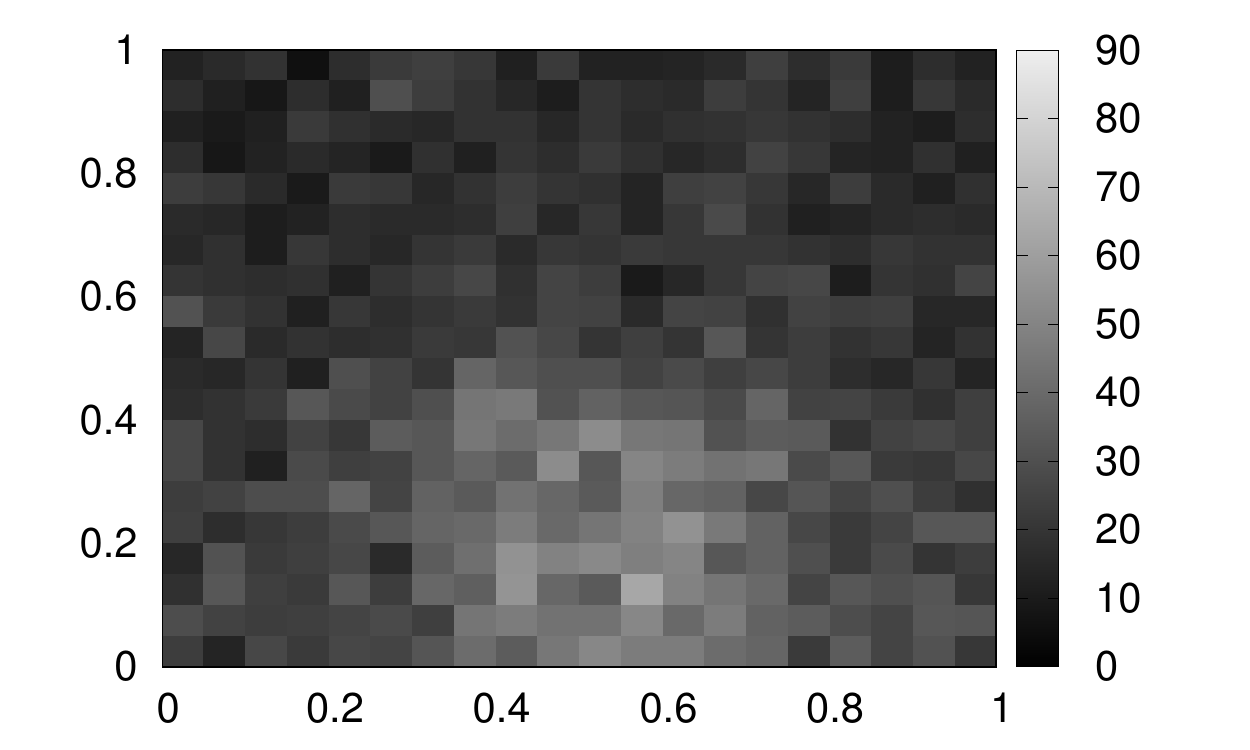}
	\scriptsize
	(c) $w_{error}=1.0,w_{size}=0.5$
	\end{minipage}
	\begin{minipage}{0.49\linewidth}
	\centering
	\includegraphics[width=\linewidth]{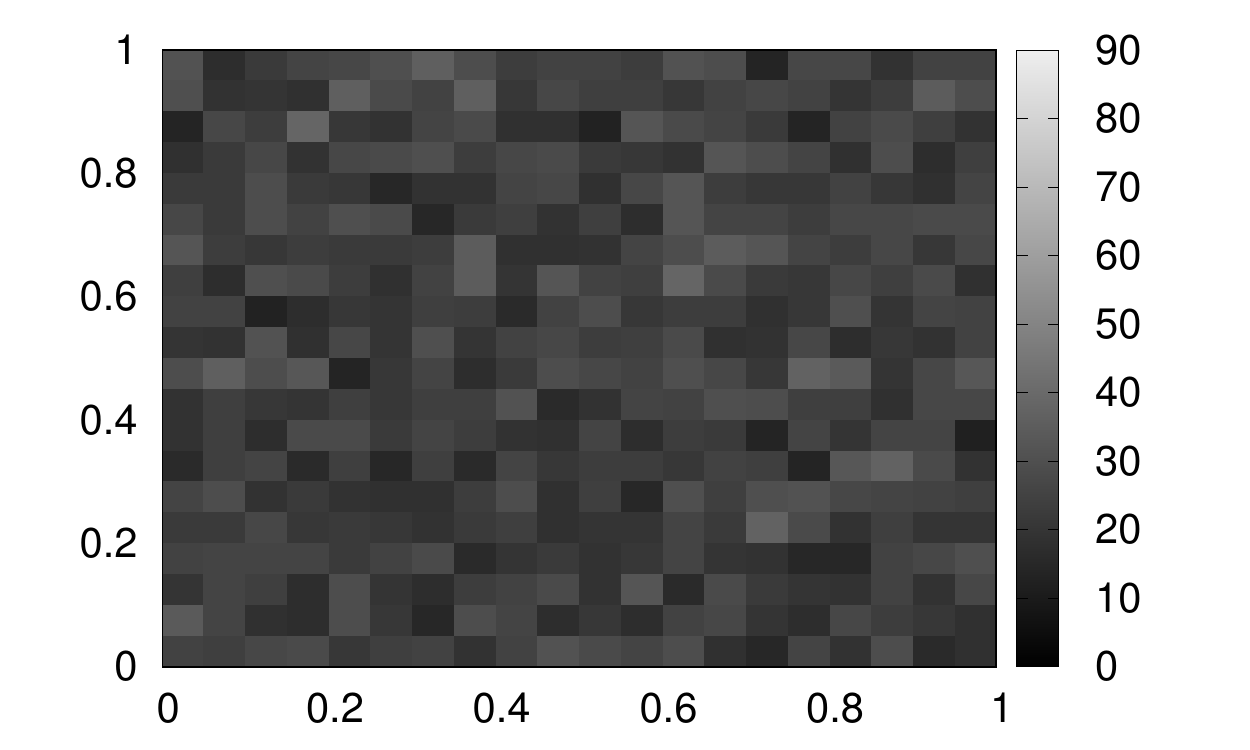}
	\scriptsize
	(d) $w_{error}=0.0, w_{size}=1.0$
	\end{minipage}
\vspace{-.3cm}
	\caption{Sample distribution for different weights}
	\label{fig:sampling:distribution}
\end{figure}
\vspace{-.3cm}

{\color{black}

To further analyze the sampler's functionality, we provide an example of execution for a known performance function of the form $y = 0.8 x_1 + 0.2x_2$. On a randomly selected point, an abnormality is introduced, modeled by a Gaussian function. In Figure \ref{fig:sampling:distribution} (a) a projection of the performance function is provided. The horizontal and vertical axes represent $x_1$ and $x_2$ respectively and the colors represent $y$ values, where the lighter colors demonstrate higher $y$ values. Algorithm \ref{algorithm:dtadaptive} is executed for different $w_{error}$, $w_{size}$ during the \texttt{SAMPLE} step. We assume a maximum number of deployments $B$ of 100 points out of the 2500 available points and a per-iteration number of deployments $b$ of 10 points. In Figures \ref{fig:sampling:distribution} (b), (c) and (d) we provide the distribution of the selected samples, for different weight values. Each dimension is divided in 20 intervals and for each execution we keep count of the samples that appear inside each region. The color of the regions demonstrate the number of the samples within the region (lighter colors imply more samples). 

The adaptiveness of the proposed methodology lead the samples to immediately identify the abnormal area. In Figure \ref{fig:sampling:distribution} (b) the score of each leaf is only determined by its error. Most samples are gathered around the Gaussian distribution: The first leaves that represent the area of the Gaussian function produce less accurate models since they cannot express the performance function with a linear model. Since the score of each leaf is only determined by its error, these leaves claim the largest share of $b$ at each step, thus the samples are gathered around the abnormality. On the contrary, when increasing $w_{size}$ as in Figure \ref{fig:sampling:distribution} (c), the gathering of the samples around the abnormality is neutralized as more samples are now distributed along the entire space, something that is intensified in Figure \ref{fig:sampling:distribution} (d), where the abnormality is no longer visible.


The consideration of two factors (error and size) for deciding the number of deployments spent at each leaf targets the trade-off of \emph{exploring} the Deployment Space versus \emph{exploiting} the obtained knowledge, i.e., focus on the abnormalities of the space and allocate more points to further examine them. This is a well-known trade-off in many fields of study \cite{sutton1998reinforcement}. In our approach, one can favor either direction by adjusting the weights of leaf error and size, respectively. It should be noted that the consideration of other parameters, such as the deployment cost, can be easily considered through the extension of the score function (Line 14). This way, more ``expensive'' deployment configurations, e.g., ones that entail multiple VMs with many cores, would be avoided in order to regulate the cost of the profiling.

}

\subsection{Modeling and Optimizations}
\label{section:profiling:modeling}

After $B$ samples are returned by the profiling algorithm, they are utilized by the \emph{CREATEMODEL} (Algorithm \ref{algorithm:dtadaptive}, Line 8) function to train a new Decision Tree. The choice of training a new Decision Tree instead of expanding the one used during the sampling phase is made to maximize the accuracy of the final model. When the first test nodes of the former Decision Tree were created, only a short portion of the samples were available to the profiling algorithm and, hence, the original Decision Tree may have initially created inaccurate partitions. To tackle this, the final set of samples is used to train the model from scratch. {\color{black} In cases where the number of obtained samples is comparable to the dimensionality of the Deployment Space, the number of constructed leaves is extremely short and the tree degenerates into a linear model that covers sizeable regions of the Deployment Space. Since many performance functions may present non-linear characteristics, the accuracy of the end model could degrade. To tackle this problem, along with the final Decision Tree, we also train a set of Machine Learning classifiers and keep the one that achieves the lowest Cross Validation error. Our evaluation demonstrated that for very low sampling rates, i.e., area (a) of Figure \ref{fig:example}, alternative classifiers such as  Perceptrons \cite{rosenblatt1961principles} and Ensembles of them \cite{dietterich2000ensemble} may achieve higher accuracy. However, when the Decision Tree is trained with enough samples, it outperforms all the other classifiers, i.e., in area (b) of Figure \ref{fig:example}. This is the main reason for choosing the Decision Tree as a base model for our scheme: The ability to provide higher expressiveness by composing multiple linear models in areas of higher unpredictability, make them a perfect choice for modeling a performance function.}

The case where the tree is poorly partitioned during the initial algorithm iterations, can also produce vulnerabilities during the sampling step. Specifically, the Deployment Space may be erroneously partitioned on the first levels of the tree and then, as new samples arrive, more deployments will be spent on creating more and shorter regions that, would otherwise have been merged into a single leaf. To address this problem, commonly anticipated when constructing a Tree in an ``online'' fashion, several solutions have been proposed \cite{tao2009efficient}. In our approach, the Decision Tree is recreated from scratch prior to the sampling step. This optimization, albeit requiring extra computation, boosts the performance of the profiling methodology, as better partitioning of the space leads to more representative regions and better positioned models. Nevertheless, the time needed for this extra step is marginal compared to the deployment time for most modern distributed applications.

\eat{
\subsection{Complexity analysis}
The most time-consuming step of our methodology is the tree expansion step (Algorithm \ref{algorithm:dtadaptive}, Line 4). Specifically, for each leaf, the algorithm requires:
$$
\mathcal{O}(\left(|u_1|\cdot|u_2|\right)^2\cdot n^2 \cdot|samples|)
$$
steps ($u_1$, $u_2$ being the value lists as expressed in Algorithm \ref{algorithm:treeexpansion}, $n$ being the dimensionality of the Deployment Space and $|samples|$ is the total number of samples), since the most time consuming part of the for loops (Algorithm \ref{algorithm:treeexpansion} Lines 9--10)  is the estimation of the linear model (Algorithm \ref{algorithm:treeexpansion}, Lines 22--23). This estimation is done for $|points|^2$ times, where $|points| = |u_1|\cdot|u_2|$, and its complexity is equal to $\mathcal{O}(n^2\cdot|samples|)$, hence the above complexity. To further simplify the above expression, we can replace the cardinality of $u_1$, $u_2$ with the highest dimension cardinality, say $k$. Since $|samples|=B$, the total complexity of the \emph{SPLITLEAF} function is equal to: $\mathcal{O}(k^4n^2B)$.
\emph{SPLITLEAF} is invoked for each leaf, the number of which is proportional to $B$, thus, the complexity of each iteration is equal to:
$$
\mathcal{O}(k^4n^2B^2)
$$
This complexity may look prohibitive, however, Algorithm \ref{algorithm:dtadaptive} is executed ``offline''. The \emph{DEPLOY} function (Line 6) dominates the execution time of the profiling algorithm in general, thus such a high complexity is marginal against the time needed to deploy the application and execute the selected workload. 
}
\sectionpadding
\section{Evaluation}
\label{section:evaluation} 
In this section, we provide the experimental evaluation of our methodology. 
First, we evaluate the performance of our scheme (Algorithm \ref{algorithm:dtadaptive}) against other end-to-end profiling approaches.
We then provide a thorough analysis of the different parameters that impact the algorithm's performance.

\subsection{Methodology and data}
To evaluate the accuracy of our profiling algorithm, we test it over various real and synthetic performance functions. We are using the \emph{Mean Squared Error} (MSE) 
metric for the comparison, estimated over the \emph{entire} Deployment Space, i.e., we exhaustively deploy all possible combinations of each application's configurations, so as to ensure that the generated model successfully approximates the original function for the entire space.  We have deployed five different popular real-world applications, summarized in Table \ref{tab:apps}.  We opted for applications with diverse characteristics with Deployment Spaces of varying dimensionalities (3 -- 7 dimensions).

The first three applications (k-means, Bayes and Wordcount) are deployed on a YARN \cite{vavilapalli2013apache} cluster. YARN creates the abstraction of a unified resource pool in terms of memory and cores. New tasks claim an amount of resources which are either provided to them (and the task is executed) or, if not available, stall until the resources become available. The Media Streaming application consists of two components: The backend is an NFS server that provides videos to the Web Servers. A number of lightweight Web Servers (nginx) are setup to serve the videos to the clients. The NFS server retains 7 different video qualities and 20 different videos per quality. 
Finally, MongoDB is deployed as a sharded cluster and it is queried using YCSB \cite{cooper2010benchmarking}. The sharded deployment of MongoDB consists of three components: (a) A configuration server that holds the cluster metadata, (b) a set of nodes that store the data (MongoD) and (c) a set of loadbalancers that act as endpoints to the clients (MongoS). Since MongoD and MongoS components are single-threaded, we only consider the cases of scaling them horizontally. Each application was deployed in a private Openstack installation, with 8 nodes aggregating 200 cores and 600GB of RAM. Due to space constraints, in Appendix B of the Extended Version of this work \cite{giannakopoulos2017decision}, a thorough application analysis is provided.

\vspace{-.5cm}
\begin{table}[htb]
\caption{Applications under profiling}
\label{tab:apps}
\footnotesize
\centering
\begin{tabular}[htb]{|l|l|c|}
\hline
\textbf{Application}& \multirow{2}{*}{\textbf{Dimensions}} & \multirow{2}{*}{\textbf{Values}}\\ 
\textbf{(perf. metric)} &  &\\ \hline
\multirow{7}{*}{Spark k-means} 
&  YARN nodes & 2--20\\
&  \# cores per node & 2--8 \\
&  memory per node & 2--8 GB \\
&  \# of tuples & 200--1000 ($\times 10^3$) \\
(execution time)&  \# of dimensions & \{1,2,3,5,10\}\\ 
&  data skewness & 5 levels\\
&  k & \{2,5,8,10,20\}\\\hline
\multirow{5}{*}{Spark Bayes} 
&  YARN nodes & 4--20\\
&  \# cores per node & 2--8 \\
&  memory per node & 2--8 GB \\
(execution time)&  \# of documents & 0.5--2.5 ($\times 10^6$) \\
&  \# of classes & 50--200 classes\\ \hline
\multirow{4}{*}{} 
&  YARN nodes  & 2--20\\
Hadoop Wordcount 
&  \# cores per node & 2--8 \\
 (execution time)&  memory per node & 2--8 GB \\
&   dataset size& 5--50 GB\\ \hline
\multirow{3}{*}{Media Streaming} 
&  \# of servers &  1--10\\
&  video quality & 144p--1440p\\
(throughput)&  request rate & 50--500 req/s\\
\hline
\multirow{3}{*}{MongoDB} &  \# of MongoD & 2--10 \\
&  \# of MongoS & 2--10\\
(throughput) &  request rate & 5--75 ($\times 10^3$) req/s\\
\hline
\end{tabular}
\end{table}
\vspace{-.3cm}


\vspace{-.3cm}
\begin{table}[htb]
\footnotesize
\caption{Synthetic performance functions}
\label{tab:functions}
\centering
\begin{tabular}[htb]{|l|l|l|l|}
\hline
\textbf{Complexity} &\textbf{Name} & \textbf{Function} & $R^2$\\ \hline
\multirow{2}{*}{LOW}
& LIN & $f_1(\mathbf{x}) = a_1 x_1 + \cdots + a_n x_n$ & $1.00$ \\
& POLY &  $f_2(\mathbf{x}) = a_1 x_1^2 + \cdots + a_n x_n^2$ & $0.95$ \\
\hline
\multirow{3}{*}{AVG}
& EXP & $f_3(\mathbf{x}) = e^{f_1(\mathbf{x})}$ & $0.65$\\
& EXPABS &  $f_4(\mathbf{x}) = e^{|f_1(\mathbf{x})|}$ & $0.62$ \\
& EXPSQ & $f_5(\mathbf{x}) = e^{-f_1(\mathbf{x})^2}$ & $0.54$ \\
\hline
\multirow{3}{*}{HIGH}
& GAUSS &  $f_6(\mathbf{x}) = e^{-f_2(\mathbf{x})}$ & $0.00$ \\
& WAVE &  $f_7(\mathbf{x}) = \cos(f_1(\mathbf{x}))\cdot f_3(\mathbf{x})$ & $0.00$\\
& HAT &  $f_8(\mathbf{x}) = f_2(\mathbf{x}) \cdot f_6(\mathbf{x})$ & $0.00$\\
\hline
\end{tabular}
\end{table}
\vspace{-.3cm}

We have also generated a set of performance functions using mathematical expressions, listed in Table \ref{tab:functions}. Each function maps the n-dimensional Deployment Space to a single dimensional metric. 
The listed functions are chosen with the intention of testing our profiling algorithm against multiple scenarios of varying complexity. To quantify each function's complexity, we measure how accurately a function can be approximated by a linear model. Linear performance functions can be approximated with only a handful of samples, hence we regard them as the least complex case. For each of the listed functions, we calculate a linear regression model that best represents the respective data points and test its accuracy using the coefficient of determination $R^2$. Values close to $1$ indicate that the specified functions tend to linearity whereas values close to $0$ indicate that the specified functions are strongly non-linear. Based on the values of $R^2$ for each case, we generate three complexity classes: Functions of \emph{LOW} complexity (when $R^2$ is higher than $0.95)$), functions of \emph{AVERAGE} complexity (when $R^2$ is between $0.5$ and $0.7$) and functions of \emph{HIGH} complexity when $R^2$ is close to $0$. The values of $R^2$ depicted in Table \ref{tab:functions} refer to two-dimensional Deployment Spaces. 
%


\twocolfivefigures{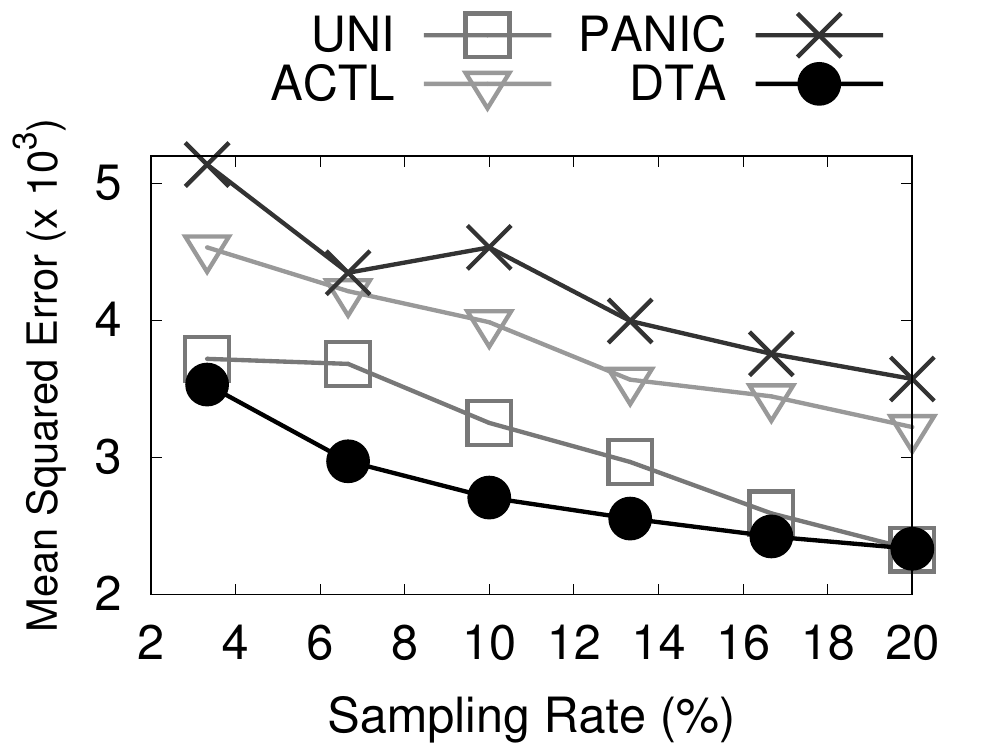}{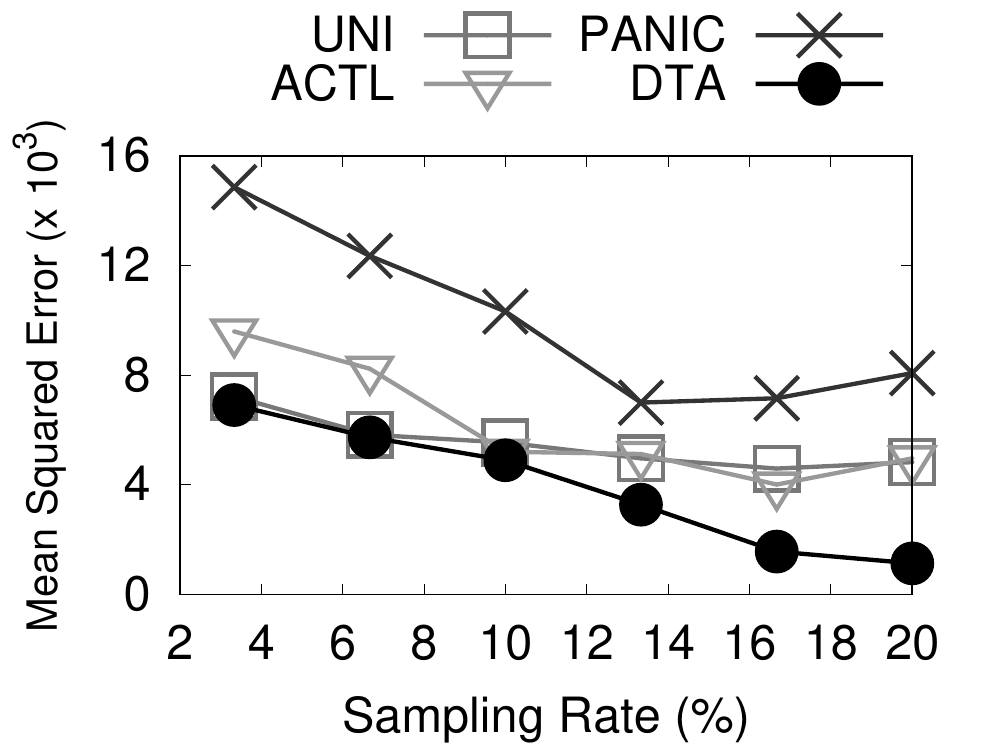}{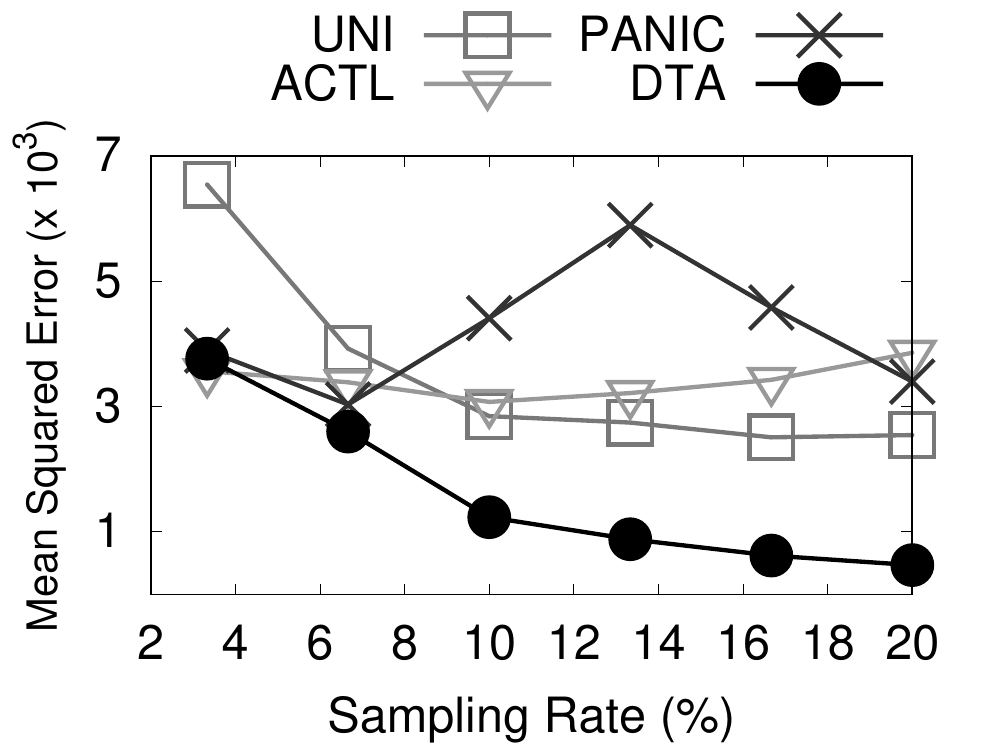}{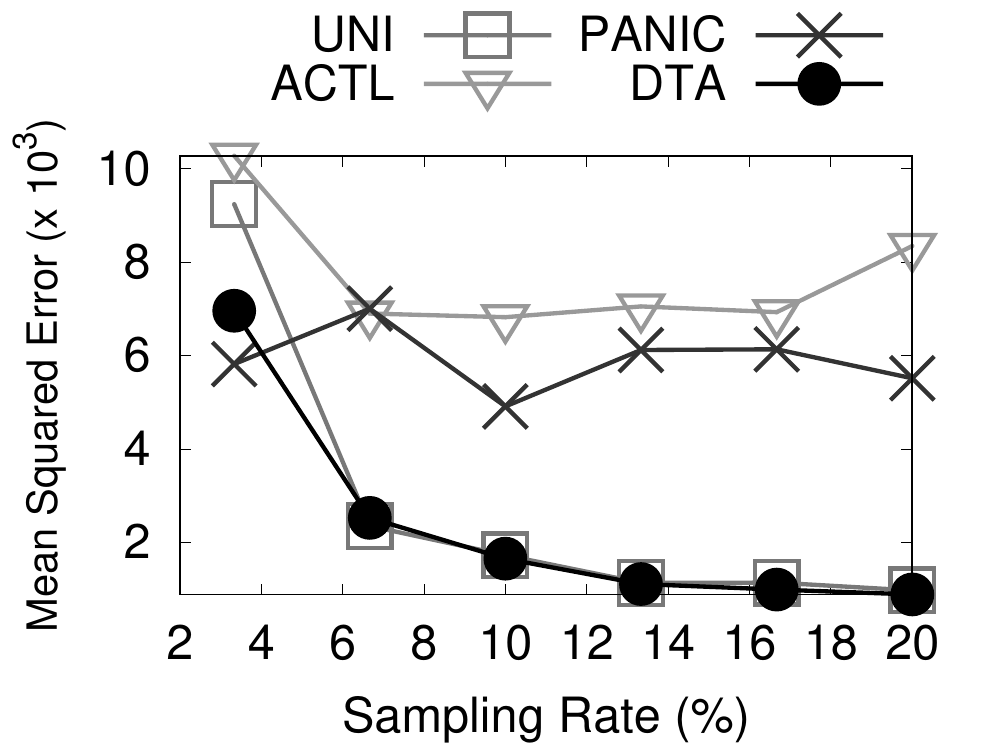}{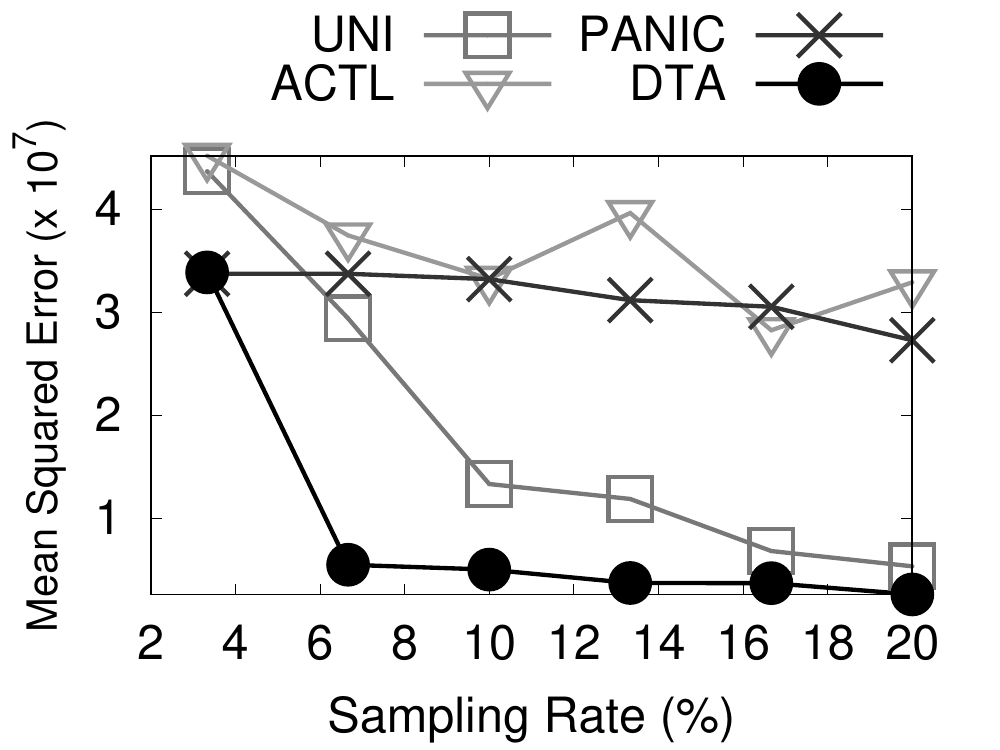}{Accuracy vs sampling rate (MSE)}{fig:4.2.1:mse}
\subsection{Profiling algorithms comparison}
\label{section:evaluation:comparison}
First, we compare our profiling methodology against other end-to-end profiling schemes. Our approach is referred to as Decision Tree-based Adaptive methodology (\emph{DTA}). { \color{black} Active Learning \cite{settles2010active} (\emph{ACTL}) is a Machine Learning field that specializes on exploring a performance space by obtaining samples assuming that finding the class or the output value of the sample is computationally hard. We implemented \emph{Uncertainty Sampling} that prioritizes the points of the Deployment Space with the highest uncertainty, i.e., points for which a Machine Learning model cannot predict their class or continuous value with high confidence. PANIC \cite{giannakopoulos2015panic} is an adaptive approach that favors points belonging into steep areas of the performance function, utilizing the assumption that the abnormalities of the performance function characterize it best.} Furthermore, since most profiling approaches use a randomized sampling algorithm \cite{gonccalves2015performance,cunha2016cloud,kundu2010application,kundu2012modeling} to sample the Deployment Space and different Machine Learning models to approximate the performance, we implement a profiling scheme where we draw random samples (\emph{UNI}) from the performance functions and approximate them using the models offered by WEKA \cite{hall2009weka}, keeping the most accurate in each case. {\color{black} In all but a few cases, the Random Committee \cite{rokach2010ensemble} algorithm prevailed, constructed using Multi-Layer Perceptron as a base classifier. 
For each of the aforementioned methodologies, we execute the experiments 20 times and present the median of the results. 

\subsubsection{Sampling rate}
We first compare the four methods against a varying Sampling Rate, i.e., the portion of the Deployment Space utilized for approximating the performance function ($SR=\frac{|D_s|}{|D|} \times 100\%$). SR varies from 3\% up to 20\% for the five real-world applications. In Figure \ref{fig:4.2.1:mse} we provide the accuracy of each approach measured in terms of MSE. 



From Figure \ref{fig:4.2.1:mse}, it becomes apparent that DTA outperforms all the competitors for increasing Sampling Rates, something indicative of its ability to distribute the available number of deployments accordingly so as to maximize the modeling accuracy. In more detail, all algorithms benefit from an increase in Sampling Rate since the error metrics rapidly degrade. Both in k-means and Bayes, when the sampling rate is around 3\% UNI and DTA construct models of the highest accuracy. 
{\color{black} 
As mentioned in Section \ref{section:profiling:modeling}, for such low Sampling Rates the linearity of the Decision Tree would fail to accurately represent the relationship between the input and the output dimensions, thus a Random Committee classifier based on Multi-Layer Perceptrons is utilized for the approximation. The same type of classifier also achieves the highest accuracy for the rest of the profiling algorithms (ACTL, PANIC) that, as seen from the Figure, present higher errors due to the less accurate sampling policy at low SR. As the Sampling Rate increases, the Decision Tree obtains more samples and creates more leaves, which contributes in the creation of more linear models that capture a shorter region of the Deployment Space and, thus, producing higher accuracy. Specifically, for $SR>3\%$, Decision Tree created a more accurate prediction than other classifiers and, hence, it was preferred.}

In the rest of the cases, DTA outperforms its competitors for the Wordcount application and, interestingly, this is intensified for increasing $SR$. Specifically, DTA manages to present $3\times$ less modeling error than UNI when $SR=20\%$. Media Streaming, on the other hand, exhibits an entirely different behavior. The selected dimensions affect the performance almost linearly, thus the produced performance function is smooth and easily modeled by less sophisticated algorithms than DTA, explaining UNI's  performance which is similar to DTA's. PANIC and ACTL try to identify the abnormalities of the space and fail to produce accurate models. Finally, for the MongoDB case, DTA outperforms the competitors increasingly with SR. In almost all cases, DTA outperforms its competitors and creates models even 3 times more accurate (for Bayes when $SR=20\%$) from the best competitor. As an endnote, the oscillations in PANIC's and ACTL's behavior are explained by the aggressive exploitation policy they implement. PANIC does not explore the Deployment Space and only follows the steep regions, whereas ACTL retains a similar policy only following the regions of uncertainty, hence the final models may become overfitted in some regions and fail to capture most patterns of the performance function. {\color{black}Our work identifies the necessity of both exploiting the regions of uncertainty but also for exploring the entire space; This compromise is a unique feature of DTA and explains its prevalence a applications difficult to approximate. Finally, notice that all the applications were evaluated for areas (a) and -- part of -- (b) of Figure \ref{fig:example}.
}

\subsubsection{Performance function complexity}
We now compare the accuracy of the profiling algorithms against the complexity of the profiling functions. We create synthetic performance functions from those presented in Table \ref{tab:functions}. For each of them, we generate two datasets with 2 and 5 dimensions, using random coefficients for each dimension and constant Deployment Space size of 10K points. We run the profiling algorithms for each function assuming sampling rates of $0.5\%$ and $2.0\%$ and present the results in Table \ref{tab:4.2.2}. Since the output dimension of each dataset is in different scale, we normalize all the results, dividing the error of each methodology with the error produced by UNI. LIN, EXP and EXPSQ were approximated successfully from all methodologies, i.e., the achieved errors were less than $10^{-5}$  so they are not included in the Table. The scores in bold demonstrate the lowest errors for each case.

{\color{black}
Table \ref{tab:4.2.2} showcases that \emph{all} synthetic functions were approximated more accurately by DTA than the rest of the profiling algorithms. Specifically, even in the most complex cases (GAUSS, WAVE and HAT), DTA achieves lower errors and the difference between the competitors is increasing with the Sampling Rate. Again, we notice that increasing SR greatly benefit the accuracy of DTA since more samples allow it to focus more on the unpredictable regions of the Deployment Space.
Regarding the rest of the profiling methodologies, the attribute of ACTL and PANIC to almost exclusively focus on regions where the performance function presents oscillating behavior leads them to an increasing difference with UNI, especially when the function's complexity increases and the oscillations become more frequent. On the contrary, when the performance functions are characterized totally from their abnormality, i.e., a discontinuity or another pattern appearing in specific regions of the Deployment Space, like in the EXPABS case, both of them create very accurate models. In this case, we observe that DTA produces results of high quality too: For both Sampling Rates it presents results similar to PANIC's and in the $2\%$ case, DTA marginally outperforms it. DTA manages to outperform all the competitors, even when the performance function is far more complex than the one of a distributed application.

\vspace{-.7cm}
\begin{table}[htb!]
\caption{Accuracy vs function complexity}
\scriptsize
\label{tab:4.2.2}
\centering
\begin{tabular}{|l|l|c|c|c|c|c|c|c|c|}
\hline
\multirow{2}{*}
{} & 
\multirow{2}{*}
{\textbf{n}} & 
\multicolumn{2}{|c|}{\textbf{ACTL}} & 
\multicolumn{2}{|c|}{\textbf{PANIC}} & 
\multicolumn{2}{|c|}{\textbf{DTA}}  
\\ 
\cline{3-8}
& & 
0.5\% & 2\% &
0.5\% & 2\% &
0.5\% & 2\% \\
\hline
\multirow{2}{*}{POLY}  
	& 2 & 1.03  & 1.57  & 1.05  & 1.23  & \textbf{0.80} & \textbf{0.89} \\ 
	& 5 & 0.99  & 1.05  & 0.98  & 1.22  & \textbf{0.97} & \textbf{0.45} \\ \hline
\multirow{2}{*}{EXPABS}
	& 2 & 8.99  & 5.29  & 1.29  & 2.63  & \textbf{0.18}  & \textbf{0.32}  \\
	& 5 & 0.40  & 0.91 & \textbf{0.13}  & 0.12  & \textbf{0.13}  & \textbf{0.11}  \\ \hline
\multirow{2}{*}{GAUSS} 
	& 2 & 1.55  & 5.91  & 0.92  & 5.63  & \textbf{0.28}  & \textbf{0.42}  \\
	& 5 & 2.25  & 5.99  & 1.86  & 1.47  & \textbf{0.99}  & \textbf{0.67}  \\ \hline
\multirow{2}{*}{WAVE} 
	& 2 & 0.98  & 2.34  & 1.01  & 2.51  & \textbf{0.91}  & \textbf{0.99}  \\
	& 5 & 1.10  & 1.13  & 1.08  & 1.02  & \textbf{0.97}  & \textbf{0.77}  \\ \hline
\multirow{2}{*}{HAT}   
	& 2 & 1.17  & 6.49  & 1.25  & 4.85  & \textbf{0.60}  & \textbf{0.42} \\
	& 5 & 8.01  & 4.77  & 6.63  & 4.20  & \textbf{0.99}  & \textbf{0.95} \\
\hline
\end{tabular}
\end{table}
\vspace{-.3cm}

To summarize, let us recall that the evaluation takes place for the \emph{entire} Deployment Space and not only for abnormal regions. As the complexity of a function increases, the ability of a profiling algorithm to identify the complex regions becomes crucial for its accuracy. Using oblique Decision Trees, DTA manages to quickly identify those regions and, as we will see in Section \ref{section:evaluation:oblique}, produce models that accurately reflect them with very small error. Furthermore, due to their inherent divide-and-conquer nature, Decision Trees are proven to be particularly efficient for Deployment Spaces of increasing dimensionality as they achieve to decompose the space according to the dimensions' importance and, as seen by Table \ref{tab:4.2.2}, retain satisfying accuracy, in contrast to the rest of the approaches. 
}

\subsection{Parameter Impact Analysis}
\label{section:evaluation:analysis}
We now study the parameters that affect DTA's performance. k-means is utilized as a test application, since it comprises the most dimensions and is the most complex among the applications we tested. In the Appendix C of \cite{giannakopoulos2017decision}, we continue our evaluation with further experiments, omitted from this version due to space constraints.

\subsubsection{Per-iteration number of deployments}
\label{sec:local_size}
One of the most important parameters in DTA's execution is $b$: This is the number of deployments distributed among the leaves of the tree in each iteration. In Figure \ref{fig:4.3.1}, we provide results where $b$ is set as a portion of $B$ (the total number of allowed deployments) for three different $SR$. The horizontal axis represents the ratio $\frac{B}{b}$. The Figure on the left depicts the MSE while the right one represents the respective execution time of DTA.

When $\frac{B}{b}=1$, the algorithm degenerates into a random algorithm such as UNI, since the whole tree is constructed in one step. At this point, the algorithm presents the highest error and the lowest execution time, since the tree is only constructed once and the most erroneous leaves are not prioritized. When the ratio increases, the algorithm produces more accurate results and the decrease in error becomes more profound as the $SR$  increases. For example, when $SR=20\%$ the error decreases more than $35\%$ while the ratio increases. However, when $SR=5\%$ and $SR=10\%$  and for low $b$ values (e.g., $\frac{B}{b}=10$) an interesting pattern becomes apparent. In these cases, the error metric starts to increase, neutralizing the effect of low $b$. This occurs in cases where $b$ is extremely small, compared to the dimensionality of the Deployment Space.
In such cases, $b$ at each iteration is harder to be evenly distributed to the leaves. The leaves with the lowest scores are almost always ignored due to rounding $b$ (since each leaf must be assigned with an integer number of samples and leaf's score is a decimal number). We can conclude that although low $b$ values have a positive effect into the algorithm's accuracy, extremely low $b$ values that make it comparable to the dimensionality of the space tend to reduce its accuracy. 
\vspace{-.3cm}
\begin{figure}[htb!]
	\centering
	\begin{minipage}{.49\linewidth}
	\centering
		\includegraphics[width=\linewidth]{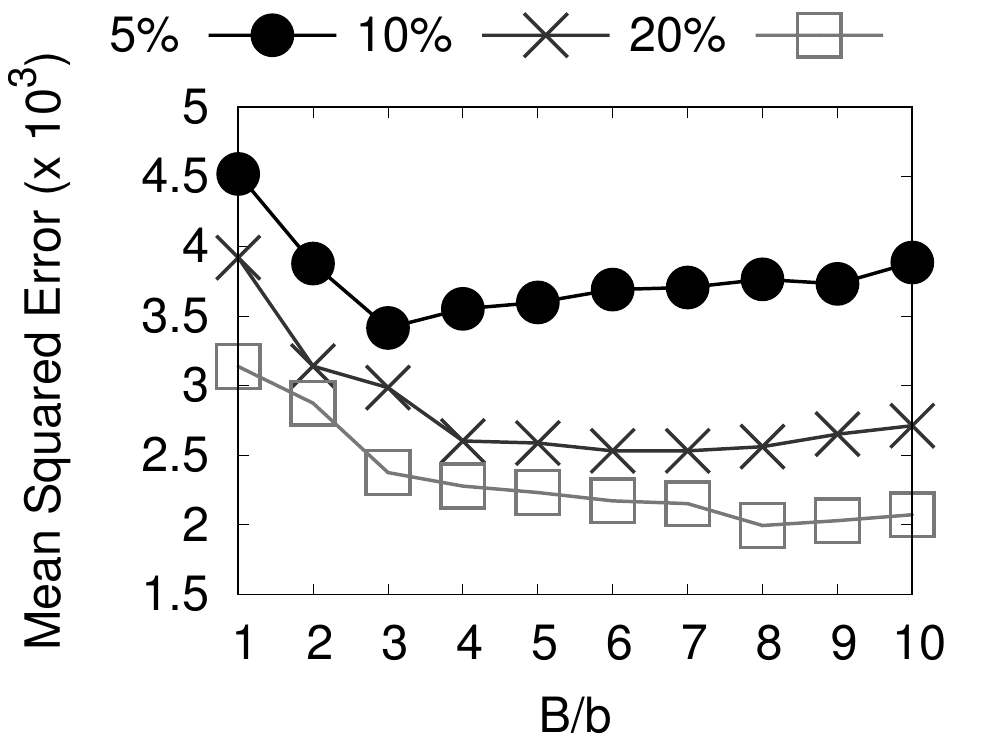}
	\end{minipage}
	\begin{minipage}{.49\linewidth}
	\centering
		\includegraphics[width=\linewidth]{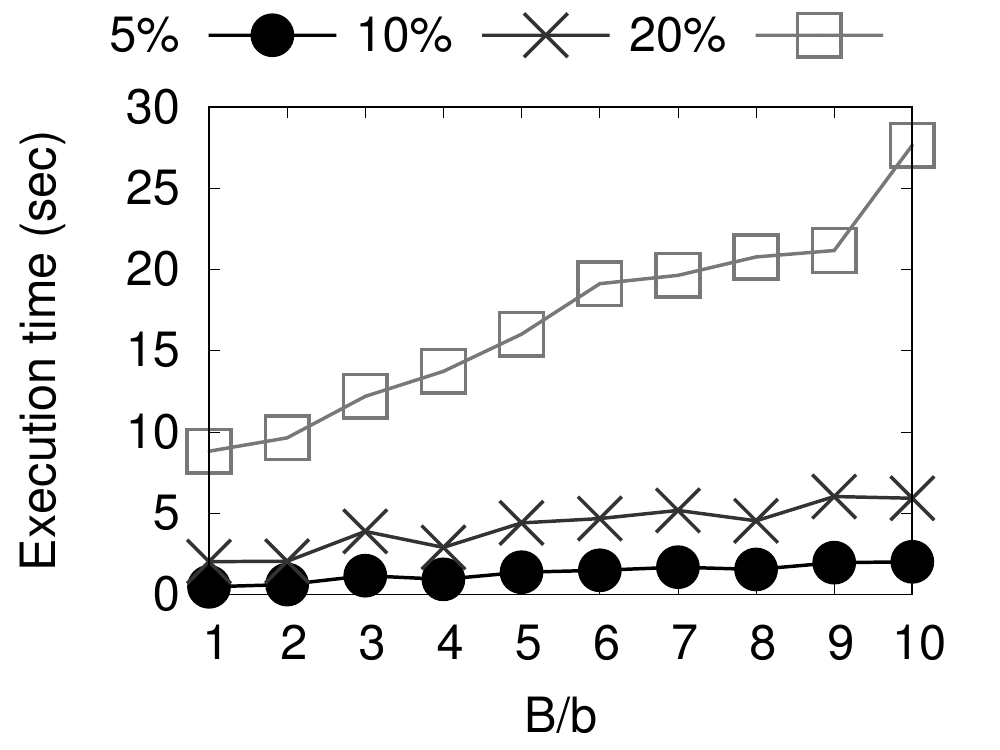}
	\end{minipage}
\vspace{-.3cm}
	\caption{Accuracy vs $B/b$}
	\label{fig:4.3.1}
\end{figure}
\vspace{-.3cm}

The performance gain observed by increasing $\frac{B}{b}$ can be attributed to two factors: (a) the final Decision Tree has more accurate cuts and, hence, well-placed models and (b) the samples are properly picked during the profiling. To isolate the impact of these two factors, we repeat the same experiment for all applications and train different ML classifiers instead of a Decision Tree. This way, we isolate the impact of sampling and examine its magnitude. We identified that the classifiers that consist of linear models, i.e., regression classifiers such as OLS (Ordinary Least Squares) \cite{dismuke2006ordinary}, or an Ensemble of Classifiers created with Bagging \cite{quinlan1996bagging} (BAG) that utilize linear models as base improve their accuracy with increasing $\frac{B}{b}$. In Table \ref{tab:sampling_models} we provide our findings for two classifiers (OLS and BAG) for four of the five applications; Media Streaming is too smooth, as it does not contain abnormalities or discontinuities, and hence a sophisticated sampling algorithm is not essential for the extraction of an accurate profile. In the Table we provide the percentage decrease in MSE compared to the case $\frac{B}{b}=1$.

The table demonstrates that for all cases, an increasing $\frac{B}{b}$ benefits the linear models, something that showcases that our sampling algorithm itself achieves better focus on the interesting Deployment Space regions. Nevertheless, we notice that an increasing $\frac{B}{b}$ ratio does not always lead to linear error reduction. When $\frac{B}{b}=8$ and $\frac{B}{b}=10$, one can notice that the error either degrades marginally (for k-means and Bayes) or slightly increases (for MongoDB) when compared to $\frac{B}{b}=5$. This is ascribed to the extremely low per-iteration number of deployments $b$: When $\frac{B}{b}$ increases, $b$ becomes comparable to the dimensionality of the Deployment Space and, hence, the proportional distribution of $b$ according to the leaves' scores is negatively affected. Other tested classifiers such as Perceptrons remain unaffected by $\frac{B}{b}$. Yet, our findings demonstrate that even utilizing solely the sampling part of our methodology can be particularly useful in cases where linear models, or a composition of them, are required.
\vspace{-.5cm}
\begin{table} [htb]
	\centering
	\caption{MSE decrease \% for linear classifiers}
	\footnotesize
	\begin{tabular}{|c|c|r|r|r|r|}
		\hline
		\multirow{2}{*}{\textbf{Application}} 
		& \multirow{2}{*}{\textbf{Classifier}} 
		& \multicolumn{4}{|c|}{\textbf{B/b}} \\ 
		\cline{3-6}
		& & \textbf{2} & \textbf{5} & \textbf{8} & \textbf{10} \\
		\hline
		\multirow{2}{*}{k-means}
		& OLS & 8\% & 10\% & 12\% & 12\% \\
		& BAG & 3\% & 8\% & 8\% & 9\% \\ 
		\hline
		\multirow{2}{*}{Bayes}
		& OLS & 23\% & 27\% & 28\% & 29\% \\
		& BAG & 23\% & 22\% & 21\% & 23\% \\ 
		\hline
		\multirow{2}{*}{Wordcount} 
		& OLS & 23\% & 17\% & 19\% & 23\% \\
		& BAG & 21\% & 28\% & 41\% & 38\% \\ 
		\hline
		\multirow{2}{*}{MongoDB}
		& OLS & 19\%& 13\% & 12\% & 7\% \\
		& BAG & 6\% & 12\% & 11\% & 2\% \\ 
		\hline
	\end{tabular}
	\label{tab:sampling_models}
\end{table}
\vspace{-.3cm}

\subsubsection{Oblique boundaries}
\label{section:evaluation:oblique}
We now compare the effect of flat versus oblique Decision Trees in the profiling accuracy and execution time. In  Figure \ref{fig:4.3.3} (a), we provide the results for the k-means application for varying $SR$.

\vspace{-.3cm}
\begin{figure}[htb!]
	\scriptsize
	\centering
	\begin{minipage}{.49\linewidth}
	\centering
		\includegraphics[width=\linewidth]{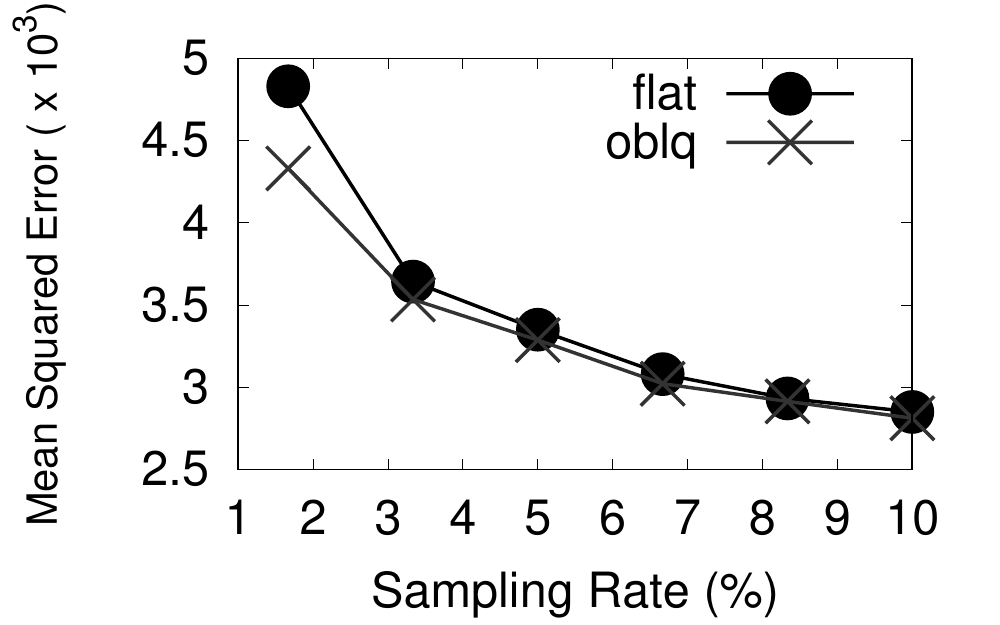}
		(a)
	\end{minipage}
	\begin{minipage}{.49\linewidth}
	\centering
		\includegraphics[width=\linewidth]{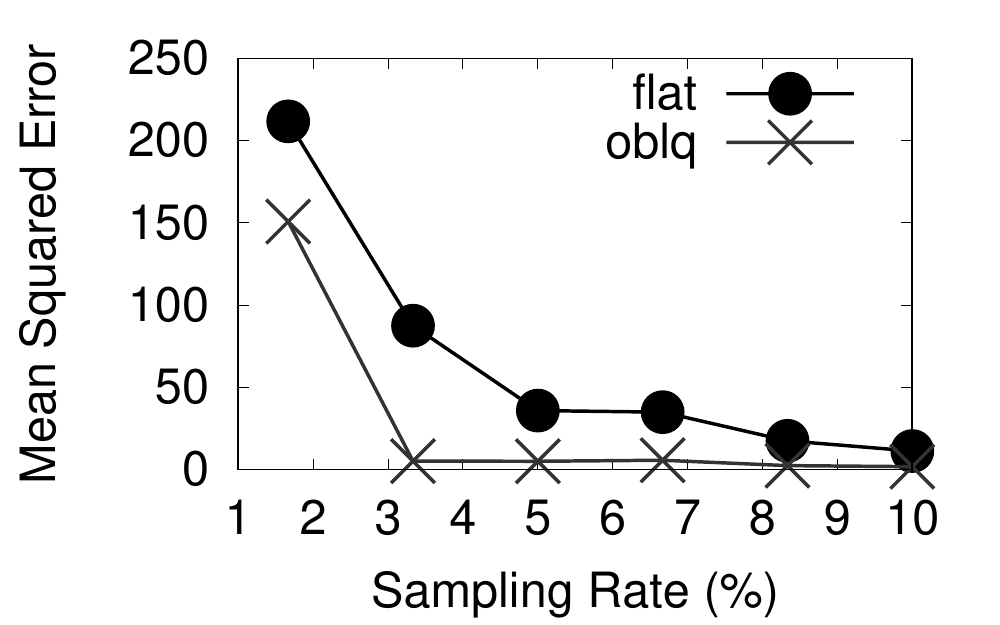}
		(b)
	\end{minipage}
	\begin{minipage}{.49\linewidth}
	\centering
		\includegraphics[width=\linewidth]{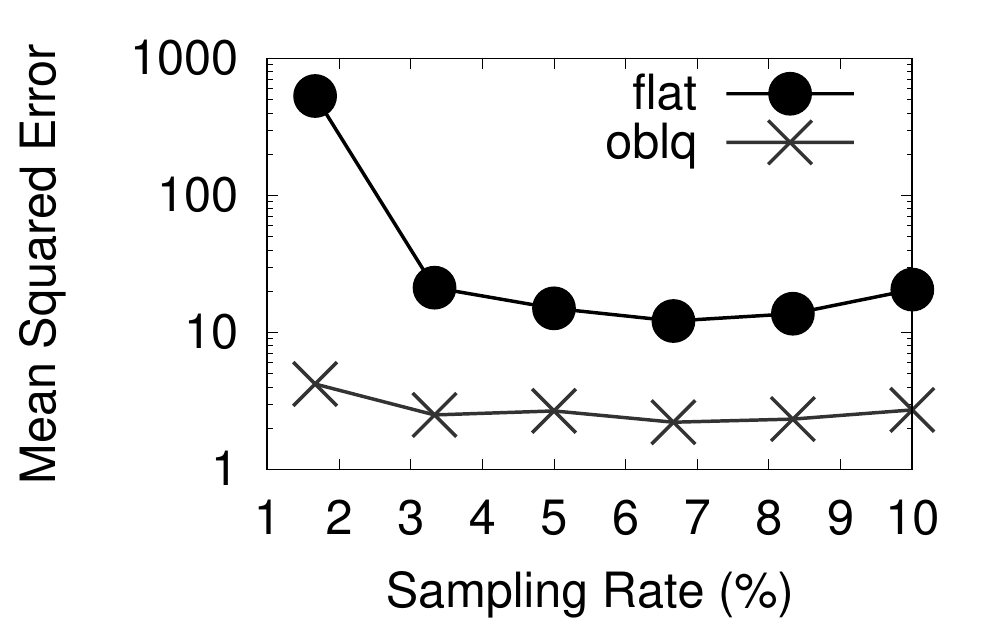}
		(c)
	\end{minipage}
	\begin{minipage}{.49\linewidth}
	\centering
		\includegraphics[width=\linewidth]{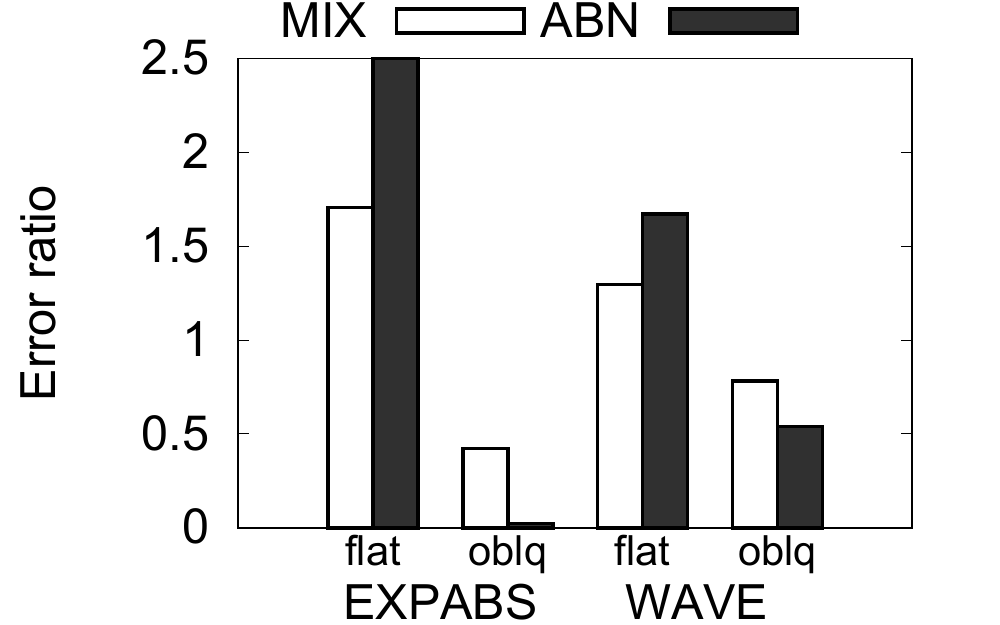}
		(d) 
	\end{minipage}
	\vspace{-.3cm}
	\caption{Accuracy vs flat and oblique cuts}
	\label{fig:4.3.3}
\end{figure}
\vspace{-.3cm}

The oblique tree introduces a slight accuracy gain (of about 10\% for low $SR$) compared to the flat tree, as seen by the left subfigure while, simultaneously, demands about twice the time for algorithm execution. The accuracy gains fade out when the $SR$ increases. This phenomenon can be attributed to the fact that, when a higher SR is employed, more leaf nodes are created and the profiling algorithm functions in a more fine-grained manner. At this point, the exact structure of the leaf nodes (if they are oblique or flat) is not very important. However, when the algorithm must work with fewer leaves, i.e., lower $SR$ , the importance of the leaf shape becomes crucial, hence the performance boost of the oblique cuts.
A point that is not shown and stressed in the results so far is that oblique Decision Trees, apart from a minor performance boost, offer the ability to accurately approximate points of the performance function with patterns spanning to multiple dimensions of the Deployment Space. When measuring the accuracy of the model in the entire Deployment Space, as in Figure \ref{fig:4.3.3} (a), this effect can be overlooked. Yet, it becomes apparent when focusing on the region containing the pattern. To showcase this, we conduct the following experiment: Assume EXPABS of Table \ref{tab:functions}, minimized  when $a_1 x_1 \cdots a_n x_n = 0$. We again compare the flat and oblique approaches for a Deployment Space formed by two dimensions but now use two different test sets: (i) points from the entire Deployment Space (as before) and (ii) points that are relatively close to the aforementioned line. Figure \ref{fig:4.3.3} (b) corresponds to the case where the test set is drawn from the entire Deployment Space, while in Figure \ref{fig:4.3.3} (c) the tests points are close to the line, i.e., less than $\epsilon=10^{-3}$ from the line. 

While the difference in (b) is considerable for the particular $SR$ (a gain of 30\% to 90\%), when testing against points close to the abnormality we observe that the oblique version manages to achieve error rates that are orders of magnitude lower than those achieved with the flat Decision Trees. To further evaluate the impact of the test set into the measured accuracy for different performance functions, we repeat the previous experiment for EXPABS and WAVE (which also contain similar complex patterns) for a $SR=2\%$. We measure the accuracy of the model when the test points are picked from (a) the entire Deployment Space (ALL), (b) close to the ``abnormal'' region (ABN) and (c) both (MIX). For MIX, half of the test points are far from the abnormality and the other half is located in a distance less than $\epsilon$. Figure \ref{fig:4.3.3} (d) showcases the respective results. The produced errors for MIX and ABN are divided with the error of ALL for each execution (ALL is thus omitted since it is always equal to 1). Our results demonstrate that, for flat Decision Trees, the modeling error increases the more we focus on the abnormal patterns of the Deployment Space. On the other hand, oblique cuts produce much more accurate results, vastly reducing their induced errors as we focus on abnormality areas. {\color{black} Different profiling algorithms were also tested around ABN: Our evaluation indicates that DTA is the only algorithm that achieves such an accuracy gain, as the other methodologies present worse results around the abnormality region. } 


\eat{

{
\color{black}
}
}
\sectionpadding
\section{Related Work}
Performance modeling is a vividly researched area. The challenge of accurately predicting the performance of a distributed application is hindered by the virtualization overhead inserted from the cloud software (hypervisors, virtual hardware, shared storage, etc.). 
The distinct approaches used to model the behavior of a given application can be graded in three categories: (a) simulation-based, (b) emulation-based and (c) approaches involving the benchmarking of the application and take a ``black-box'' view. In the first case, the approaches are based on known models of the cloud platforms \cite{calheiros2011cloudsim} and enhance them with known performance models of cloud applications. CDOSim \cite{fittkau2012cdosim} is an approach that targets to model the Cloud Deployment Options (CDOs) and simulate the cost and performance of an application. CloudAnalyst \cite{wickremasinghe2010cloudanalyst} is a similar work that simulates large distributed applications and studies their performance for different cloud configurations. Finally, WebProphet \cite{li2010webprophet} is a work that specializes in web applications. All of these works assume that performance models regarding both the infrastructure and the application are known. On the contrary, our approach makes no assumptions neither for the application nor for the provider.

To bypass the assumption of a known performance model, emulation has been used. The idea is to deploy the application and capture performance traces for various scenarios. The traces are then ``replayed'' to the infrastructure in order to to predict the application performance. CloudProphet \cite{li2011cloudprophet} is an approach used for migrating an application into the cloud. It collects traces from the application running locally and replays them into the cloud, predicting the performance it should achieve over the cloud infrastructure. //Trace \cite{mesnier2007trace} is an approach specializing in predicting the I/O behavior of a parallel application, identifying the causality between I/O patterns among different nodes. In \cite{hoste2006performance} a similar approach is presented, in which a set of benchmark applications are executed in a cloud infrastructure, measuring microarchitecture-independent characteristics and evaluating the relationship between a target and the benchmarked application. According to this relationship, a performance prediction is extracted. Finally, in \cite{mytilinis2015performance}, a performance prediction approach is presented specialized in I/O-bound applications. Through microbenchmarking, a performance model for the virtualized storage is extracted and applied to I/O intensive BigData applications, so as to predict their performance. Although emulation approaches can be extremely efficient for capturing specific aspects of an application's behavior, they cannot be generally applied if the application structure is unknown.

Several works try to overcome this limitation and propose methodologies that make no assumption regarding the application structure. Such approaches assume that the application is a black-box that receives a number of inputs (deployment configurations) and produces a number of outputs which correspond to application performance metrics. The application is deployed for some representative deployment configurations and performance metrics are obtained. The model is then constructed by utilizing statistical and machine learning techniques mapping the configuration space into the application performance. In \cite{gonccalves2015performance,cunha2016cloud} a generic methodology is described used to infer the application performance of based on representative deployments of the configuration space. The approach tackles the problem of generalizing the performance for the entire deployment space, but does not tackle the problem of picking the most appropriate samples from the deployment space, as the suggested approach. PANIC \cite{giannakopoulos2015panic} is a similar work, that addresses the problem of picking representative points during sampling. This approach favors the points that belong to the most steep regions of the Deployment Space, based on the idea that these regions characterize most appropriately the entire performance function. However it is too focused on the abnormalities of the Deployment Space and the proposed approach outperforms it. Similarly, the problem of picking representative samples of the Deployment Samples is also addressed by Active Learning \cite{settles2010active}. This theoretical model introduces the term of uncertainty for a classifier that, simply put, expresses its confidence to label a specific sample of the Deployment Space. Active Learning favors the regions of the Deployment Space that present the highest uncertainty and, as PANIC, fail to accurately approximate the performance function for the entire space, as also indicated by our experimental evaluation.  Finally, in \cite{kundu2010application} and \cite{kundu2012modeling} two more generic black-box approaches are provided, utilizing different machine learning models for the approximation. Neither of these works, though, address the problem of picking the appropriate samples, since they are more focused on the modeling problem. 


\sectionpadding
\vspace{-.3cm}
\section{Conclusions}
\vspace{-.3cm}
In this work, we revisited the problem of performance modeling for applications deployed over cloud infrastructures. Their configuration space can grow exponentially large, making the required number of deployments for good accuracy prohibitively large. We proposed a methodology that utilizes oblique Decision Trees to recursively partition and sample the  Deployment Space, {\color{black}assuming a maximum number of cloud deployments constraint}. Our approach manages to adaptively focus on areas where the model fails to accurately approximate application performance, achieving superior accuracy under a small number of deployments. We demonstrated that our method better approximates both real-life and synthetic performance functions.
\sectionpadding

\clearpage
\bibliographystyle{abbrv}
\balance
\bibliography{bibliography}

\appendix
\section{System overview}
\label{appendix:system_overview}
As showcased in Algorithm \ref{algorithm:dtadaptive}, application profiling entails: (a) The sampling of the Deployment Space, (b) the Deployment of the chosen samples and (c) the construction of the end model. So far, we have focused our analysis and evaluation on the theoretical aspects of the methodology, ignoring the engineering part that entails the automatic deployment of the application, the enforcement of different resource- and application-level configuration parameters and the retrieval of the performance metric. Although these aspects are orthogonal to our methodology, in the sense that different cloud infrastructures, solutions and software can be utilized without altering the proposed profiling algorithm, the utilized solutions can have a great impact in the profiling time and the level of automation. In this section, we provide a brief overview of the tools utilized to create a fully-automated application profile. 

Our implementation was based on Openstack and its deployment tool, Heat \cite{heat}. Heat is responsible for the deployment and the orchestration of virtual resources, including Virtual Machines, virtual networks, ports, block devices, volumes, etc. An application description, referred to as a Heat Orchestration Template (\emph{HOT}), is, essentially, the blueprint of the application, describing its architecture and main components and supports a set of parameters that can be defined during its instantiation. This way, Heat decouples the static application information, e.g., its architecture, from the dynamic information that must be provided in runtime, e.g., the flavors of the different VMs, application level parameters, etc. 

When a new application deployment is spawned, Heat first allocates the necessary resources, in the correct order. For example, if a VM depends on a volume, the VM allocation starts upon the volume creation. After all the VMs are launched, Heat executes user-specified scripts, included in the HOT, in order to orchestrate the resources. Furthermore, to ease the application configuration, Heat also supports parameterized script execution that enables the scripts to be executed with different parameters and ensure that different resources are properly utilized. For example, the number of cores or the amount of RAM of a VM can be provided as a parameter during the HOT instantiation and, according to the value of this parameter, the script can set the appropriate parameter in the application's configuration file. Through this powerful mechanism, the problem of describing a complex application is reduced into creating a parameterized Heat template that can be launched according to user-defined parameters. 

In our case, we created four HOTs for each application of Table \ref{tab:apps} and mapped each Deployment Space dimension into a unique HOT parameter (per application). This way, the deployment of a single Deployment Space point entails the deployment of the respective HOT with the appropriate parameters that reflect the aforementioned point. The workload-specific parameters are addressed in a unified way, as HOT parameters can also be visible to the user-specified scripts. When the workload finishes, the performance metrics are inserted into a centralized database and processed in order to resume the profiling algorithm execution. To reduce the deployment time, we have constructed precooked VM images (based on Ubuntu x64 14.04 images) that contain the necessary software. This way, the deployment only entailed the resource allocation, VM bootstrapping and workload execution. 

In our experimental platform, the average VM creation time is around $60\pm20$ sec and the average VM bootstrap time is $30\pm10$ sec. Both times are measured to be independent of the VM flavor, i.e., the amount of cores, RAM and disk utilized by the VM. The workload execution time greatly varies among different applications and, in our case, dominates the entire deployment time. For each application, we estimate the average execution time along with the respective standard deviation and present them in Table \ref{tab:app_workload}.
\begin{table}[htb]
	\centering
	\caption{Workload execution times (sec)}
	\label{tab:app_workload}
	\begin{tabular}{|l|c|c|}
		\hline
		\textbf{Application} & \textbf{Average Time} & \textbf{Std. Deviation} \\ 
		\hline
		k-means& 120 & 300 \\
		Bayes & 250 & 250 \\
		Wordcount & 270 & 250 \\
		Media Streaming & 350 & 350\\
		MongoDB & 150 & 130 \\
		\hline
		
	\end{tabular}
\end{table}

It must be noted that our approach is capable of parallelizing the deployment of different samples: When the expansion of the Decision Tree in a specified step is finished, the samples that have been chosen can be deployed in-parallel. We recall here that the amount of samples chosen at each algorithm step is determined by $b$, as described in Section \ref{section:profiling}. When $b$ becomes equal to $B$, that corresponds to the maximum number of cloud deployments executed by our algorithm, our approach degenerates into a Random Sampler and the highest parallelization is achieved, although accuracy is sacrificed, as shown in Section \ref{sec:local_size}. Taking into consideration this comment and based on the average Deployment and Workload execution times, we can roughly estimate the average time needed for the estimation of a profile for application $A$, as a function of $B$ and $b$ as follows: 
$$
ProfilingTime(B,b,A) = \left \lceil{\frac{B}{b}} \right \rceil \times (T_{creation}+T_{boot}+T_{w}(A))
$$
where $T_{creation}$ refers to the VM creation time, $T_{boot}$ refers to the VM boot time and $T_{w}$ is the workload execution time for application $A$. 

\section{Application analysis}
\color{black}
\label{appendix:apps}
In this section we provide a thorough analysis of the applications utilized throughout the experimental evaluation, in Section \ref{section:evaluation}. We choose to demonstrate our profiling methodology over applications with diverse characteristics covering both \emph{batch} and \emph{online} functionality. k-means, Bayes and Wordcount are typical examples of the former category, while Media Streaming and MongoDB are representative online ones. We choose two batch applications implemented in two different systems (Spark and Hadoop respectively) in order to cover both in-memory and disk-based computations respectively. Regarding the online applications, Media Streaming was chosen for two reasons: First, media/video streaming applications are very popular and tend to become an excellent use case for cloud computing \cite{wowza,streambox,brightcove}, especially due to its decentralized and on-demand resource allocation nature. Second, its architecture, consisting of a storage backend and a set of horizontally scalable Web Servers, is very popular among cloud applications \cite{chee2010cloud}. Finally, MongoDB is a popular document store, widely used as a database backend among various applications. In all cases, we opt for representative use cases of applications deployed over cloud infrastructures. Nevertheless, we emphasize that both the nature and the behavior of the selected application is orthogonal to our approach, as the proposed methodology successfully models applications of varying diversity.

\begin{table}[htb]
	\caption{Application complexity and correlations}
	\label{tab:correlation}
	\centering
	\begin{tabular}{|l|l|c|c|}
		\hline
		\textbf{Application} & \textbf{$R^2$} & \textbf{Dimension} & \textbf{$r$} \\	\hline
\multirow{7}{*}{\textbf{Spark k-means} }
&&  YARN nodes & -0.30 \\
&&  \# cores per node & -0.35 \\
&&  memory per node & -0.32\\
&0.65&  \# of tuples &  0.66\\
(execution time)&&  \# of dimensions & 0.87\\ 
&&  data skewness & 0.14\\
&&  k & 0.45\\\hline
		\multirow{5}{*}{\textbf{Bayes}}
			&&  YARN nodes 	& -0.24\\
			&& \# cores/node 	& -0.34 \\
			&$0.56$& memory/node 	& -0.49 \\
(execution time)	&& \# of documents 	& 0.51  \\
			&& \# of classes 	& 0.00\\ \hline
			\multirow{4}{*}{} 
			&& YARN nodes  	& -0.44\\
\textbf{Wordcount}	&$0.57$& \# cores/node 	& -0.20\\
(execution time)	&& memory/node 	& -0.38\\
			&&  dataset size	& 0.49\\ \hline		
		\multirow{3}{*}{\textbf{Media Streaming}} 
			&& \# of servers 	& 0.46\\
			&$0.46$& video quality 	& -0.41\\
(throughput)		&&   request rate	& 0.28\\ \hline
		\multirow{3}{*}{\textbf{MongoDB}} 
			&&  \# of MongoD 	& 0.33\\
			&$0.55$&  \# of MongoS 	& 0.50\\
(throughput)		&& request rate 	& 0.44\\ \hline
	\end{tabular}
\end{table}

In order to provide more information regarding the behavior of the chosen applications, we first evaluate the nature of the performance functions and compare them to a linear function with the methodology described in Section \ref{section:evaluation}: Using \emph{all} the available performance points, we construct a linear model that best fits the data (with Ordinary Least Squares) and, then, compare this model to the original data points using $R^2$. In Table \ref{tab:correlation} we provide our findings. According to the categorization provided in Table \ref{tab:functions} and based on the applications' scores, we observe that all the tested applications belong in the  AVERAGE category and are quite different from their linear equivalent functions. 

The selected dimensions for each application were chosen in order to reflect an actual profiling scenario, in which the application dimensions can vary in importance. To quantify this importance, we measure the correlation between the input and output dimensions using Pearson correlation coefficient $r$, as demonstrated in Table \ref{tab:correlation}; The output dimensions are listed beneath the application names.
A negative sign indicates that the growth of one dimension has a negative impact on the output dimension. For example, when increasing the number of YARN nodes in Bayes, the execution time of the job decreases. Values close to zero indicate that the specified dimensions are of low importance. For example, in Bayes, the number of classes used for the classification does not affect the execution time of a job. It is important for a profiling methodology to be able to efficiently handle dimensions that are, eventually, proven to be of low importance, since it is quite possible that when a new application is submitted for profiling, the user cannot be aware of which dimensions have a greater impact on the application performance. Such input dimensions increase the dimensionality of the Deployment Space, without adding any extra information. Our scheme manages to efficiently handle these dimensions due the utilization of Decision Trees: The multidimensional cuts, created during the sampling phase, ignore the least important dimensions and focus on splitting the performance function based on the more important variables, increasing this way the profiling accuracy.

\section{Extended evaluation}
\subsection{Cost-aware profiling}
\label{section:evaluation:cost}
So far, we have assumed that all Deployment Space points are equivalent, in the sense that we have not examined the deployment configuration they represent: A point representing a deployment configuration of 8 VMs, each of which has 8 cores, is equivalent to a point representing 1 VM with 1 core. However, since the choice of a point results in its actual deployment, it is obvious that this choice implicitly includes a (monetary) cost dimension that has not been addressed. We now examine DTA's ability to adapt when such a cost consideration exists. Let us define the following cost models:
\begin{myitemize}
\item Bayes: $|$nodes$|\times |$cores$|$
\item Wordcount: $|$nodes$|\times |$cores$|$
\item Media Streaming: $|$servers$|$
\item MongoDB: $|$MongoS$|$ + $|$MongoD$|$
\end{myitemize}
We have chosen realistic cost models, expressed as functions of the allocated resources (VMs and cores). Let us recall that Media Streaming and MongoDB utilize unicore VMs. Hence, their cost is only proportional to the number of allocated VMs.

%
We execute DTA, utilizing the following score function for each leaf:
\begin{equation}
score(l) = w_{error} \cdot error(l) + w_{size} \cdot size(l) - w_{cost} \cdot cost(l)
\label{eq:score}
\end{equation}
in which the monetary cost of the points of each leaf are penalized, according to the pricing scheme of the provider.

For $w_{error}=1.0$ and $w_{cost}=0.5$, we alter the weight of the cost parameter between $0.2$ and $1.0$ for SR of $3\%$ and $20\%$. We provide our findings in Table \ref{tab:sampling}, in which we present the percentage difference in the profiling error (measured in MSE) and cost for each case, against the case of $w_{cost}=0.0$.

\begin{table} [htb]
	\footnotesize
	\centering
	\caption{MSE and Cost for different cost weights}
	\def\arraystretch{0.9}
	\setlength\tabcolsep{.13cm}
	\begin{tabular}{|c|c|r|r|r|r|r|r|}
		\hline
		\multirow{2}{*}{\textbf{App/tions}} 
		& \multirow{2}{*}{\textbf{SR}} 
		& \multicolumn{3}{|c|}{\textbf{MSE}} 
		& \multicolumn{3}{|c|}{\textbf{Cost}} \\ 
		\cline{3-8}
		&
		& \textbf{0.2} & \textbf{0.5} & \textbf{1.0} 
		& \textbf{0.2} & \textbf{0.5} & \textbf{1.0} \\
		\hline
		\multirow{2}{*}{\textbf{Bayes}} 
		& 3\% & +1\% & -1\% & -2\% & -1\% & -1\% & -1\%\\
		& 20\% & +4\% & +10\% & +5\% & -7\% & -9\% & -12\%\\ 
		\hline
		\multirow{2}{*}{\textbf{Wordcount}} 
		& 3\% & -1\% & -5\% & -1\% &-4\% & -6\%& -1\%\\
		& 20\% & 0\% & +13\% & +19\% & -6\% & -8\% &-18\%\\ 
		\hline
		\multirow{2}{*}{\textbf{Media Str.}}
		& 3\% & -1\% & -3\% & +3\% & -1\% & -5\% & -6\%\\
		& 20\% & -6\% & -2\% & -8\% & -7\% & -11\% & -26\%\\ 
		\hline
		\multirow{2}{*}{\textbf{MongoDB}}
		& 3\% & +6\% & +12\% & +13\% & -2\% & -3\% & -4\%\\
		& 20\% & -1\% & -7\% & -7\% & -6\% & -9\% & -12\%\\ 
		\hline
	\end{tabular}
	\label{tab:sampling}
\end{table}

For low sampling rates, increasing values for $w_{cost}$ does not heavily influence the profiling cost. Specifically, for the MongoDB case, the cost reduces by a marginal factor (around $4\%$ in the most extreme case) whereas the error increases by $13\%$. On the contrary, for high sampling rates, it becomes apparent that the consideration of the cost increases its impact as the application profiles are calculated even $26\%$ less expensive  than the case of $w_{cost}=0$, e.g., in the Media Streaming case for $w_{cost}=1.0$. Furthermore, it is obvious that for increasing $w_{cost}$, the cost becomes a more important factor for the leaf score and, hence, the cost degradation becomes more intense. Regarding the profiling accuracy, in most cases the error remains the same or its increase does not exceed $10\%$. A notable exception from this is the Wordcount case, where we can see that the MSE increases rapidly with increasing $w_{cost}$ values and even reaches a growth of $19\%$ in the case where $w_{cost}=1.0$. From this analysis, we can conclude that cost-aware sampling becomes particularly effective for high sampling rates, which is also desirable since high sampling rates entail many deployments, i.e., increased cost. In such cases, the cost-aware algorithm has more room to improve the profiling cost whereas, on the same time, the accuracy sacrifice is totally dependent on the nature of the performance function; However, from our evaluation we can conclude that the accuracy degradation is analogous to the cost reduction, allowing the user to choose between higher accuracy or reduced deployment cost.

\label{appendix:eval}
\subsection{MAE for varying Sampling Rate}
\label{appendix:mae}
Figure \ref{fig:4.2.1:mae} describes the accuracy of the proposed methodology against UNI, PANIC and ACTL for varying Sampling Rates, measured in terms of MAE.
\onecolfigures{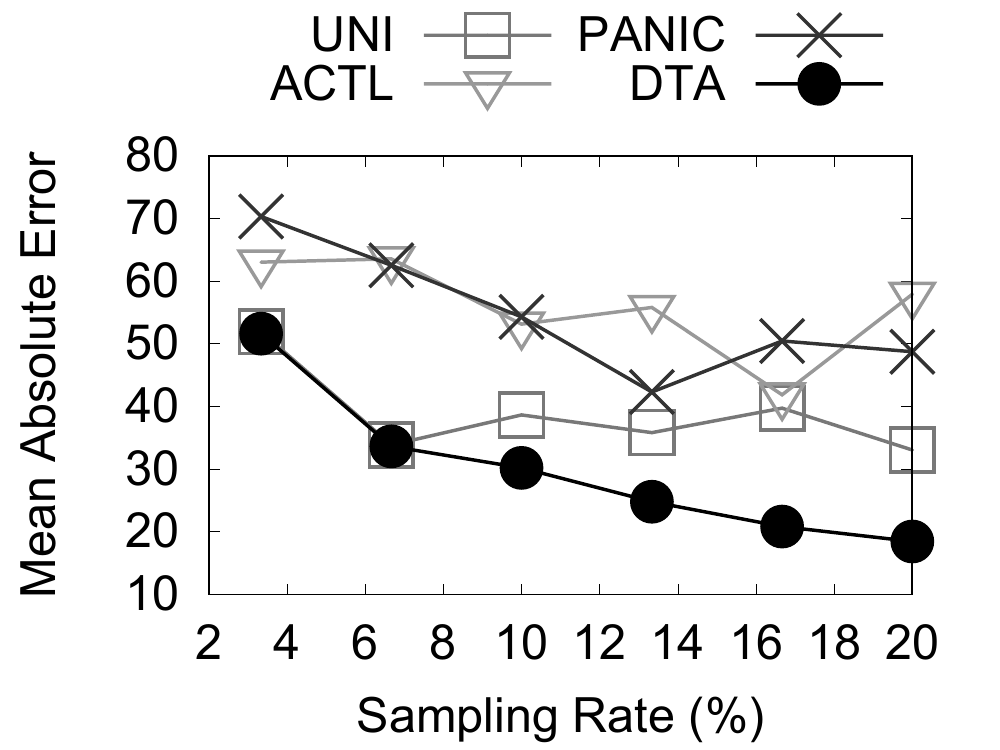}{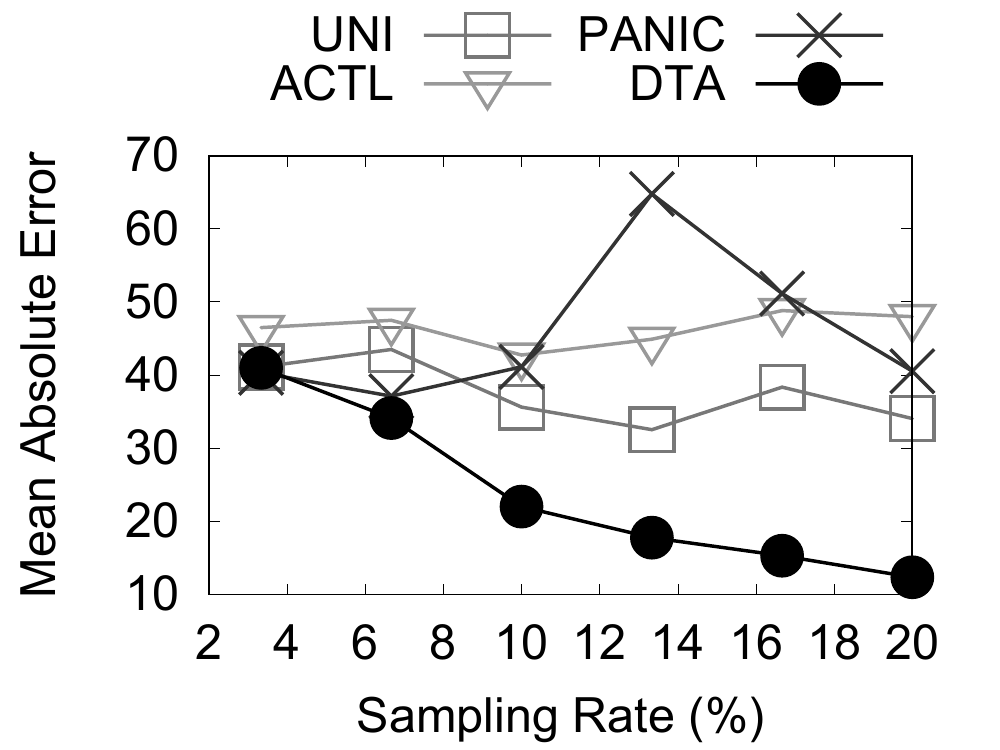}{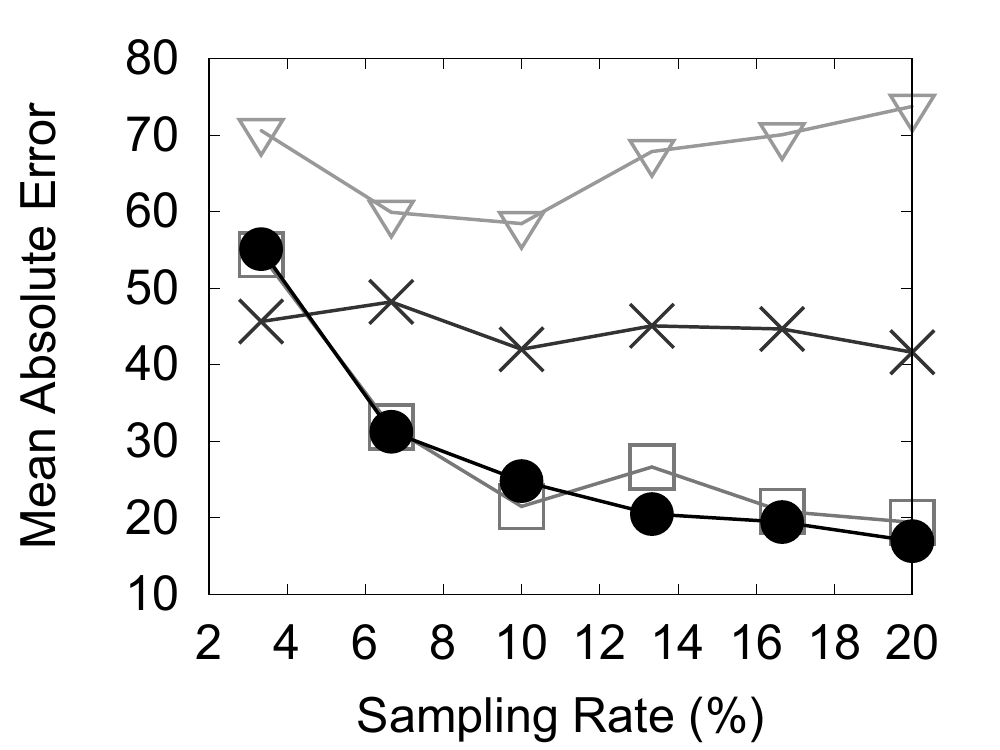}{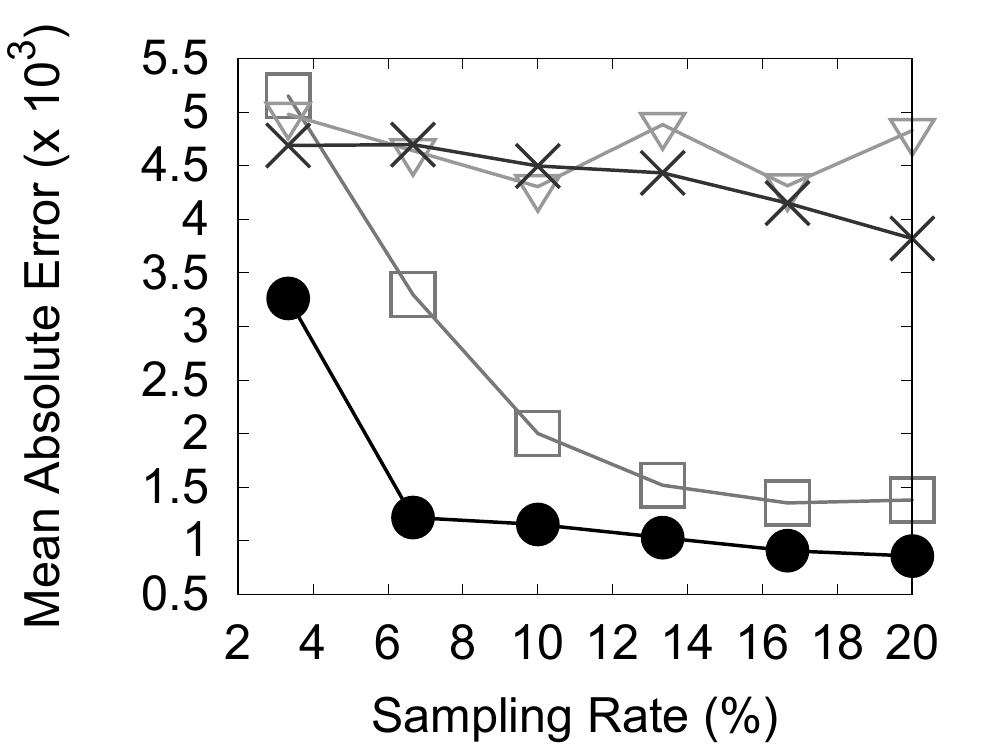}{Accuracy vs sampling rate (MAE)}{fig:4.2.1:mae}{0cm}

Comparing Figure \ref{fig:4.2.1:mae} with Figure \ref{fig:4.2.1:mse}, it is apparent that both MSE and MAE provide a similar picture between the different profiling methodologies as DTA outperforms its competitors in all cases, UNI seems to present the next best results and PANIC and ACTL present the worst accuracy, due to their aggressive exploitation policy that focuses on the abnormalities of the Deployment Space. Specifically, DTA produces results even $3.5$ times better than the next best algorithm (for MongoDB, $SR=6\%$) and the gap between the profiling methodologies increases with increasing $SR$ for all cases, but for the Media Streaming case, where UNI achieves the same accuracy level.

\subsection{Optimizations}
We evaluate the impact of the optimization presented in Section \ref{section:profiling:modeling}, i.e., the construction (from scratch) of the Decision Tree, so as to correct erroneous cuts of the Deployment Space during the initial algorithm iterations. In Figure \ref{fig:4.3.4}, we provide results for varying $\frac{B}{b}$, $SR=10\%$ for the case of Bayes. The \emph{online} training scheme represents the incremental expansion of the Decision Tree at each iteration whereas the \emph{offline} training scheme refers to the case of reconstructing the tree at each iteration.
\begin{figure}[htb!]
	\centering
	\begin{minipage}{.49\linewidth}
	\centering
		\includegraphics[width=\linewidth]{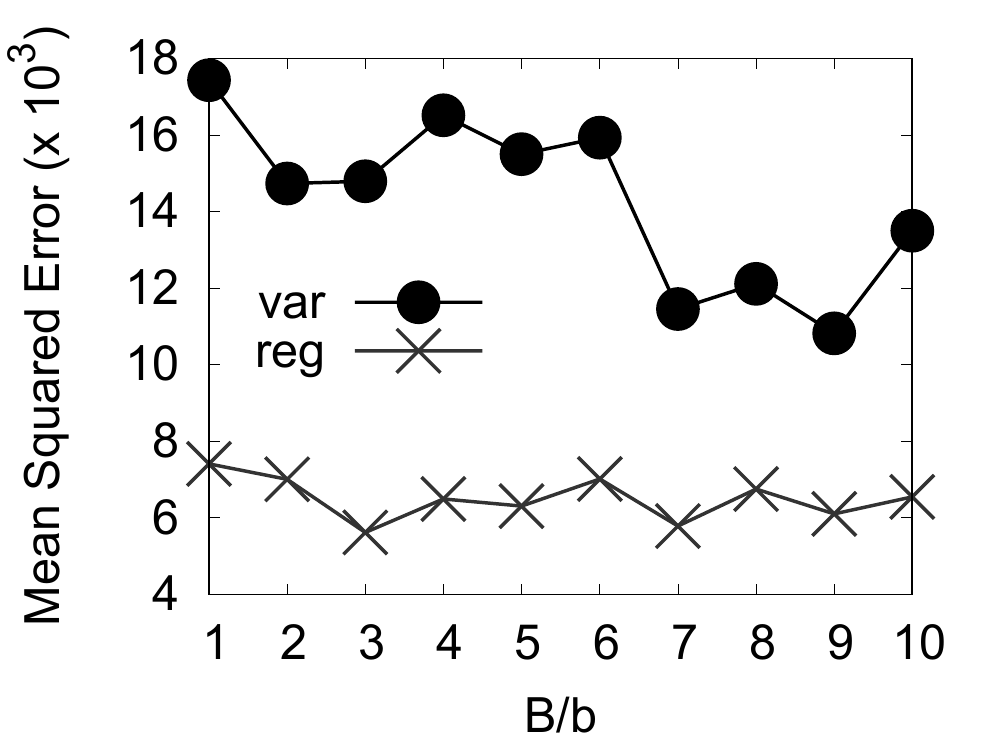}
	\end{minipage}
	\begin{minipage}{.49\linewidth}
	\centering
		\includegraphics[width=\linewidth]{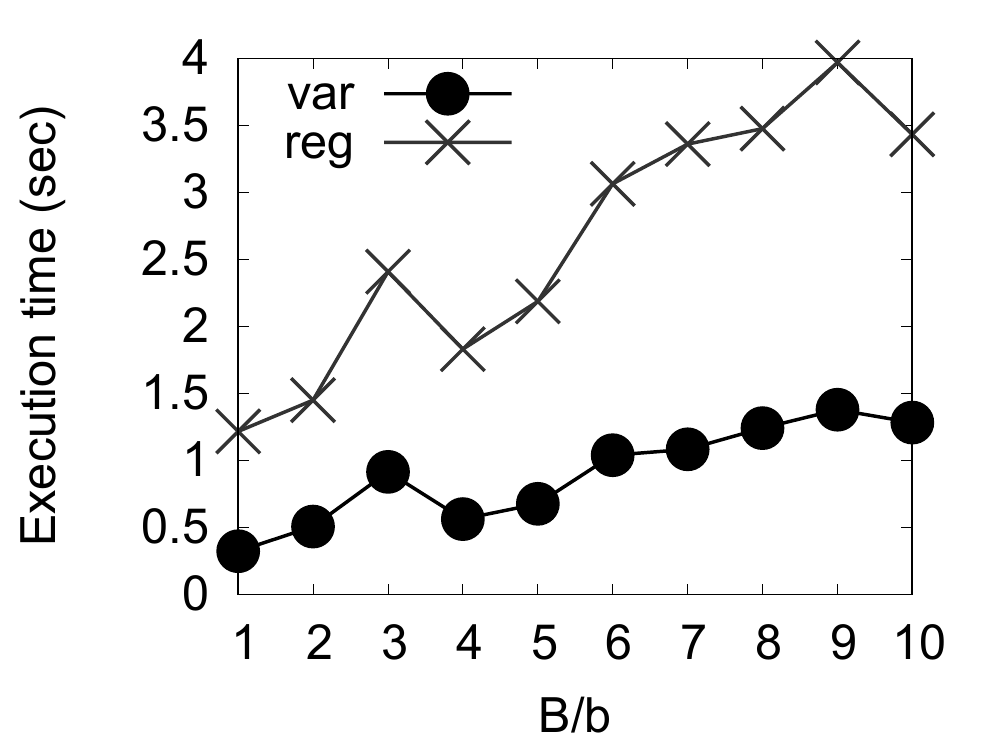}
	\end{minipage}
	\caption{Accuracy vs training type}
	\label{fig:4.3.4}
\end{figure}

When $b$ receives relatively high values (compared to $B$), the impact of offline training is marginal to the performance function. While $\frac{B}{b}$ increases, the impact of offline training increases, since the profiling algorithm produces models with lower error. This behavior is reasonable, since small values of $b$ lead to more algorithm iterations, meaning that the cuts made on the initial steps of the algorithm were decided based on a very small portion of the samples. This leads to the conclusion that offline training boosts the algorithm's accuracy when $\frac{B}{b}$ receives high values or, equivalently, $b$ is minimized. In terms of execution time, since offline training entails the total reconstruction of the Decision Tree, the algorithm needs considerably more time to terminate, as demonstrated by the right plot in Figure \ref{fig:4.3.4}. However, since the time needed to deploy and orchestrate the application is considerably larger, an overhead in the order of a few seconds is marginal. 

\subsection{Partitioning}
\label{appendix:partitioning}
As presented in Section \ref{section:profiling:construction}, a leaf node is separated by a boundary line in such a way that the samples of the different groups best fit into linear models. In this Section, this partitioning mechanism is compared to the most popular Variance Reduction methodology, that targets to minimize the intra-group variance of the target dimension. In Figure \ref{fig:4.3.2}, we provide DTA's error and execution time for the two different partitioning techniques, for different $\frac{B}{b}$, $SR=5\%$ for the Bayes application. 

From Figure \ref{fig:4.3.2}, it becomes apparent that the Variance metric fails to partition the leaves as accurately as the regression methodology. Specifically, the regression technique achieves MSEs of $1.5$ -- $2.5$ times lower than of the variance technique. The extra computation cost increases the execution time of the algorithm from $1.5$ to $4$ seconds, a difference which is marginal to the deployment cost of the application. We have to note here that the rest of the homogeneity heuristics, such as GINI impurity,  Information Gain, etc., were not considered since they are only used for classification. 

\begin{figure}[htb]
	\centering
	\begin{minipage}{.49\linewidth}
	\centering
		\includegraphics[width=\linewidth]{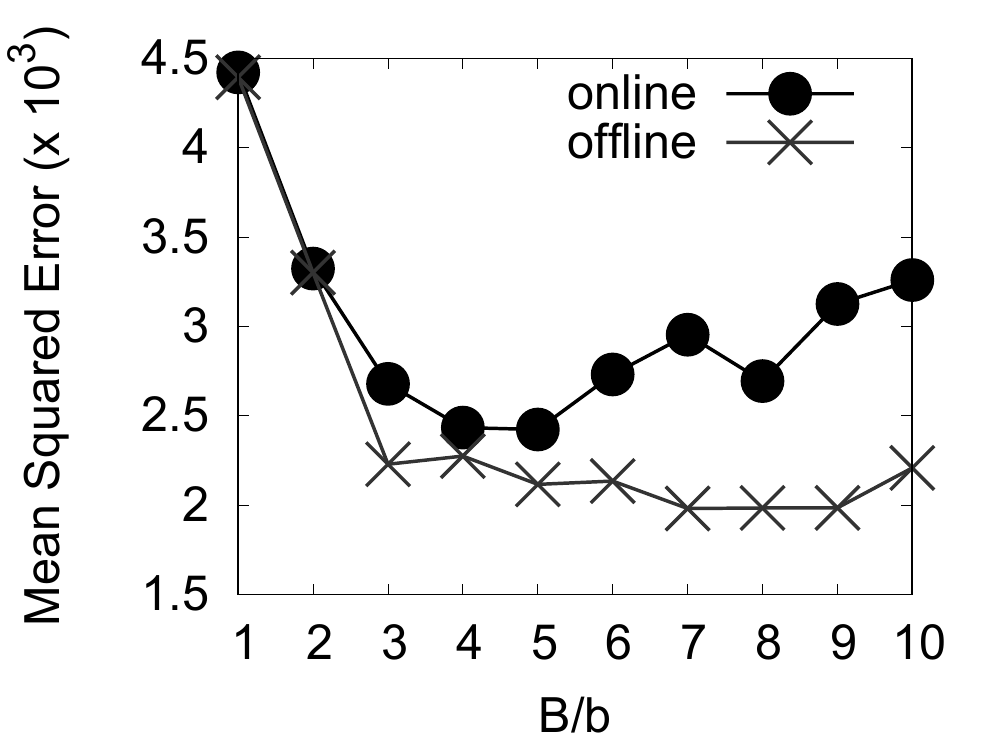}
	\end{minipage}
	\begin{minipage}{.49\linewidth}
	\centering
		\includegraphics[width=\linewidth]{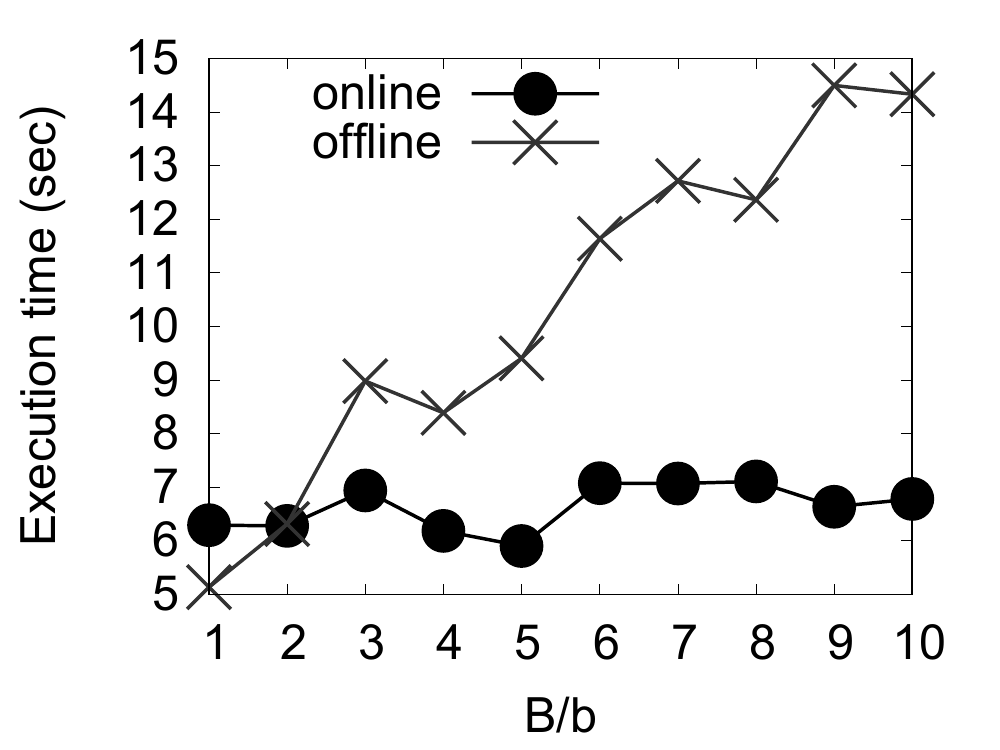}
	\end{minipage}
	\caption{Accuracy vs partitioner type}
	\label{fig:4.3.2}
\end{figure}

\end{document}